\documentclass[a4paper,fleqn]{cas-sc}

\usepackage[authoryear,longnamesfirst]{natbib}
\usepackage{algorithm}
\usepackage{algpseudocode}
\usepackage{blindtext}
\usepackage{amssymb}
\usepackage{balance}
\usepackage{amsmath}
\usepackage{booktabs}
\usepackage{soul}
\usepackage[T1]{fontenc}
\usepackage{float}
\usepackage{hyperref}
\usepackage{enumitem}

\def\tsc#1{\csdef{#1}{\textsc{\lowercase{#1}}\xspace}}
\tsc{WGM}
\tsc{QE}
\tsc{EP}
\tsc{PMS}
\tsc{BEC}
\tsc{DE}

\begin{document}
\let\WriteBookmarks\relax
\def\floatpagepagefraction{1}
\def\textpagefraction{.001}

\shorttitle{}

\shortauthors{Mohammad Zia~Ur~Rehman et~al.}

\title [mode = title]{User-Aware Multilingual Abusive Content Detection in Social Media}                      



%
\author[1]{Mohammad Zia~Ur~Rehman}[type=editor,
                        auid=000,bioid=1,
                        prefix=,
                        role=,
                        orcid=0000-0001-6374-8102]

\cormark[1]


\ead{phd2101201005@iiti.ac.in}


\credit{Conceptualization, Methodology, Software, Investigation,
Writing - Original Draft}

\affiliation[1]{organization={Department of Computer Science and Engineering},
    addressline={Indian Institute of Technology Indore}, 
    country={India}}

\author[1]{Somya Mehta}
\ead{cse190001058@iiti.ac.in}
\credit{Software, Formal analysis, Data Curation, Investigation, Writing - Original Draft}

\author[1]{Kuldeep Singh}
\ead{cse190001030@iiti.ac.in}
\credit{Software, Formal analysis, Investigation, Visualization, Writing - Original Draft}

\author[2]{Kunal Kaushik}
\ead{193083@nith.ac.in}
\credit{Software, Formal analysis, Data Curation, Investigation, Writing - Original Draft}

\author[1]{Nagendra Kumar}[type=editor,
                        auid=000,bioid=1,
                        prefix=,
                        role=,
                        orcid=]
\ead{nagendra@iiti.ac.in}

\credit{Conceptualization, Methodology, Supervision}

\affiliation[2]{organization={National Institute of Technology Hamirpur},
    country={India}}



\cortext[cor1]{Corresponding author}



\begin{abstract}
Despite growing efforts to halt distasteful content on social media, multilingualism has added a new dimension to this problem. The scarcity of resources makes the challenge even greater when it comes to low-resource languages. This work focuses on providing a novel method for abusive content detection in multiple low-resource Indic languages.
Our observation indicates that a post's tendency to attract abusive comments, as well as features such as user history and social context, significantly aid in the detection of abusive content.
The proposed method first learns social and text context features in two separate modules. The integrated representation from these modules is learnt and used for the final prediction. 
To evaluate the performance of our method against different classical and state-of-the-art methods, we have performed extensive experiments on SCIDN and MACI datasets consisting of 1.5M and 665K multilingual comments respectively. 
Our proposed method outperforms state-of-the-art baseline methods with an average increase of 4.08\% and 9.52\% in F1-scores on SCIDN and MACI datasets respectively.
\end{abstract}




\maketitle

\section{Introduction}

\label{sec:sa_int}
The rapid growth in the number of internet and social media users has brought about significant changes in the way people interact with each other and access information. It has also created new opportunities for individuals to connect with people of different cultures. However, it has also led to concerns about privacy, security, and the potential for online harassment and misinformation. A worrying trend visible in social media posts these days is to abuse an individual or a group based on their nationality, ethnicity, religion, or sexual orientation~\citep{sharma2022ceasing}.
Social media was developed for innovative and positive uses, yet a lot of abusive content in different languages is frequently posted. 
Such materials may have adverse psychological impacts including increased mental stress and emotional outbursts. 
Most social media platforms have implemented strict rules to restrict abusive content. Despite these limitations, a large amount of abusive content bypasses these impositions and is easily disseminated on social media.  
With the aim of curbing the dissemination of objectionable content, academicians, policymakers, and stakeholders are working towards developing reliable computational systems. However, there are still several challenges that need to be tackled~\citep{akiwowo2020proceedings}, \citep{roy2020framework}.
Complexities which rise due to multilingualism and code-mixed content are the major reasons for undetected abusive content. The objective of this research is to deal with following challenges that exist in detecting abusive content on social media:\\

\newcommand{\smallHead}[1]{%
    \par\noindent{#1}%
    \par\noindent\ignorespaces%
}
\smallHead{\textbf{Spelling Inconsistencies Resulting from Roman Script for Multilingual Content}}
Due to the multitude of languages spoken globally, it is common for individuals to communicate in multiple languages on social media. Some of these languages are low-resource, which further complicates the challenging task of abuse identification.
Languages with a relatively limited amount of data available for training are referred to as low-resource languages, with many Asian and African languages falling under this category.
Indic languages such as Hindi, Tamil, Marathi are a few examples of low-resource languages. Many of the West-European languages such as English, French, Spanish, Italian are high resource languages with English being the most well resourced language. The practice of writing content in native low-resource languages using Roman script is widespread today. Transcribing a native language using Roman alphabets results in several textual inconsistencies\footnote{ K. Guha, Why writing hindi in roman rather than devanagri would be a disaster (2015).}.
For instance, the Romanized Hindi word \textit{humara} can be transcribed in multiple spelling structures as \textit{humaraa, hamara}, with all forms being acceptable. More number of spelling structures are seen in the case of abusive words. Since the use of abusive words is not very frequent, users tend to write abusive words as per their own understanding. These multiple spelling structures of words hamper the performance of abusive content detection methods as they fail to recognize each occurrence of these words. Thus, it is essential to develop an efficient technique to detect the proliferation of abusive content in Romanized multilingual text.\\

\smallHead{\textbf{Different Sentence Structure in Code-Mixed Language}}
Most social media platforms have made it convenient for users to write their content in complex code-mixed text, that is people post their content using words of more than one language. For instance, the sentence \textit{Main lunch break me call karunga tumhe}, which translates to English as \textit{I will call you during the lunch break}, is a Hindi-English code-mixed sentence. This code-mixed sentence uses English words \textit{lunch break} and \textit{call} in a Hindi structured sentence. Code-mixed content is vastly used for low-resource languages, but the grammatical structure used for code-mixing may vary. The grammatical structure can be either of English or of the native language depending on the user's taste~\citep{mathur2018did}. The above mentioned code-mixed sentence follows Hindi grammatical structure, whereas the sentence, \textit{Mere maa-papa are the best in the world} which translates to English as \textit{My parents are the best in world}, follows English grammatical structure. Both of the above mentioned sentences are Hindi-English code-mixed sentences that follow different grammatical structures. 
The intricate nature of such cases presents a challenge in recognizing and controlling inappropriate content in low-resource code-mixed texts.\\

\smallHead{\textbf{Greater Emphasis on Text-based Features}}
With the emergence of advanced text feature extraction techniques like transformer-based methods, many researchers have shifted their focus solely on textual data, disregarding other valuable features. Even though text is the primary feature, social context features such as likes and reports can also provide useful insights about the nature of the content. To citep an instance of the same, it can be observed that abusive content tends to receive more reports and fewer likes. Such insights can be utilized alongside textual modality. Moreover, incorporation of users' history and disposition of posts can be beneficial for abusive content detection. Disposition of a post or post polarity refers to the distribution of number of abusive and non-abusive comments on the post. Earlier works do not capitalize on the collective benefits of all the aforementioned perspectives for abusive content detection.\\

\smallHead{\textbf{Extensive Usage of Multilingualism among Social Media Users in India}}
Our motivation to work with Indic languages can be attributed to the vast multilingualism in India. As per the 2011 census, there are a total of 1,369 rationalised mother tongues in India, that are spoken by 10,000 or more speakers, and they are further grouped into a total of 121 languages, out of which 22 are officially recognized by the constitution of India \footnote{C. of India, Census Data 2001 : General Note,[Online; accessed 19-July-2001] (2001).
URL https://censusindia.gov.in/census.website/data/census-tables/}.  Also, India has the second highest number of internet as well as social media users in the world. Recently, WhatsApp banned over 2.6 million accounts in India through its abuse detection system\footnote{WhatsApp bans over 37 lakh accounts in India in November — telegraphindia.com, https://www.telegraphindia.com/business/
whatsapp-bans-37-16-lakh-accounts-in-india-in-november/cid/1905109,
[Accessed 22-Dec-2022].}, which shows the massive amount of abusive content present on the internet, a lot of which is propagated in Indic languages. In light of the above, it has become essential to focus on a robust method to detect abusive content in Indic languages.

\begin{table}[h!]
\caption{Examples of comments}
 \centering
\begin{tabular}{llll}
\toprule
\textbf{}& \textbf{Comments}    &\textbf{Translation}              & \textbf{Label}             \\     \midrule
 1 & Ninkkaḷ eppaṭi irukkirirkaḷ...\textbf{all good}? &  \textit{How are you...all good?}               & \multirow{3}{*}{Non-abusive}
 \\
2 & \textbf{Busy} hu main, \textbf{work load} zyada hai. &  \textit{I am busy due to heavy work load}                               &                              \\
3 & Ye sab kya hai? samajh nhi aaya. & \textit{What is this? could not understand.}    &                              \\ \midrule
 1 & \textbf{Dont comment here}, anaathai kaluthai. &   \textit{Dont comment here, Orphaned Do**key.}      & \multirow{3}{*}{Abusive}     \\
2 & Iska \textbf{face} to \textbf{dog} jaisa hai. &   \textit{His face looks like a dog }                                    &                              \\
3 & Kya ch***pa hai ye!!!  & \textit{What is this nonsense!!} &                              \\ \bottomrule
\end{tabular}
\label{table:comments_example}
\end{table}

Table-\ref{table:comments_example} demonstrates examples of abusive and non-abusive content in Indic languages. First example in both categories demonstrates Inter-sentential code-mixing in Tamil-English where two sentences in a comment are written in two different languages. Second example in both categories represents Intra-sentential code-mixing in Hindi-English. Third example is written only in Hindi in both categories. To this end, we have proposed a cross-lingually trained transformer-based method to tackle these cases by extracting contextual embeddings from user generated content. The following points summarize our key contributions for the above mentioned challenges:
\begin{enumerate}
\item{
The suggested method leverages both social context and textual features by first training them in two separate modules to create a high-level representation of the features. It is shown that combining the outputs of these modules results in a significant improvement in performance.}
\item{We propose a novel multilingual abuse identification framework for multiple Indic languages. Unlike earlier techniques which focus either on monolingual corpora of regional languages~\citep{velankar2023mono,waseem2016hateful}, or bilingual corpus of code-mixed language~\citep{chopra2020hindi}, our method achieves state-of-the-art results for a total of 12 major Indic languages.} 
\item{User-generated content on social media usually includes a lot of words which do not add significantly to the information carried by such content. It is imperative to filter out these words during data preprocessing. With this in consideration, we have created a collection of 2,250 words for Indic languages.} 
\item{Another challenge is dealing with spelling errors in user-generated content. Instead of costly operation of focusing on all spellings errors, our method focuses only on spelling errors of abusive words.  
To deal with this, we have constructed a comprehensive set of abusive words for Indic languages. Set contains more than 17K abusive words including multiple spelling structures for every single abusive word. This set has been used for data augmentation, and for assisting in calculation of user and post polarity pertaining to abuse.}
\item{To deal with the code-mixed aspect of the content, we employ a cross-lingual method for text embedding extraction. Further, multiple embeddings are used in the ensemble framework to exploit the combined benefits of transformer-based methods.}
\item{We have conducted extensive experiments on two datasets to assess the efficacy of the proposed approach. 
The experimental results demonstrate that our work outperforms state-of-the-art methods by incorporating information from text as well as social context features.}
\end{enumerate}

\section{Related Work}
\label{sec:sa_rw}

 In this section, we present previous works related to abusive and hateful content detection on social media. In existing methods, the problem is framed as a supervised text classification task~\citep{schmidt2017survey}. To address the problems of text classification, there are generally two decisions to make, first is the selection of the text feature extraction technique, and next is the selection of the classification method, i.e., Machine Learning or Deep Learning methods. We present a review on existing literature based on these two categories. Further, we give an overview of existing works on Indic and multilingual methods pertaining to abusive/hateful content detection.
\subsection{\textit{Text Feature Extraction Methods}}  
Text feature extraction methods include classical Natural Language Processing (NLP) techniques for feature construction that provide  context-independent surface features, and the more recent paradigm of transformer-based methods which automatically learn abstract features from raw text.
\subsubsection{\textit{Classical Methods}}
Classical methods of text feature extraction include rule based approach, manual feature designing, count-based and frequency-based methods such as Bag of words (BoW) \& TF-IDF, and non-contextual word-embeddings.
Earliest works use rule-based approach and manually designed features extensively. 
Rule-based methods may provide good results for a set of data, but these methods are not scalable. 
N-grams and BoW methods are comparatively easier to scale. Greevy \textit{et al.}~\citep{greevy2004classifying} compares the performance of BoW, Bi-grams and part of speech (POS) tags with  Support Vector Machine (SVM) as the classifier. In their experiments BoW performs better than the other two feature extraction techniques. Xiang \textit{et al.}~\citep{xiang2012detecting} uses topical features extracted using Latent Dirichlet Allocation (LDA) algorithm. They also take lexicon features through keyword matching for vulgar language. In Davidson \textit{et al.}~\citep{davidson2017automated}, TF-IDF is used as text feature extraction technique along with POS tags and sentiment lexicons. SVM performs better than other Machine Learning methods in this work. Works like, ~\citep{waseem2016hateful, kwok2013locate}  and~\citep{waseem2016you} also use classical approaches extensively with Machine Learning methods. In the aforementioned works, a major problem is the sparsity of feature vectors. Word-embeddings overcome this problem by providing dense feature vectors of fixed length which have relatively smaller dimensionality. Glove~\citep{pennington2014glove} and Word2vec~\citep{mikolov2013efficient} are the two famous early word-embedding techniques. In~\citep{kapoor2019mind, park2017one} and~\citep{pitsilis2018effective}, word-embeddings are used as text features for hate classification. The drawback of these word-embeddings is that they are non-contextual, which hampers the performance of the hate detection systems.
\subsubsection{\textit{Transformer-based Methods }}
To overcome the drawbacks of classical methods, techniques have been developed to generate contextual embeddings. 
The major upgrade in NLP techniques is attributed to Google. Vaswani \textit{et al.}~\citep{vaswani2017attention}, at Google brain, introduces Transformer which uses attention mechanism. Since their introduction in 2017, transformers and attention mechanism have been used in a variety of NLP methods, with Bidirectional Encoder Representations from Transformers (BERT)~\citep{devlin2019bert} being the most famous method among those. 
Ranasinghe \textit{et al.}~\citep{ranasinghe2020multilingual} uses cross-lingual embeddings, generated by the transformer-based Cross-lingual Language Model-RoBerta (XLM-R), for offensive language identification. It achieves an F1-score of 85.80\%, significantly higher than its counterparts. 
Mozafari \textit{et al.} \citep{mozafari2019bert} and Chakravarthi \textit{et al.} \citep{chakravarthi2020corpus} use transformer-based methods for hateful and offensive content detection. Few works suggested language-specific methods for embedding generation such as Nguyen \textit{et al.}~\citep{nguyen2020phobert}, which provides embeddings for the Vietnamese language.
One advantage of transformer-based model is the contextual embeddings, another major advantage is transfer-learning ,i.e., sharing and re-utilising the model weights developed during training \citep{kumar2022mucot}. If training is done on a corpus for a language with a substantial resource base, its knowledge can help the process of training for low-resource languages that are still being researched~\citep{ranasinghe2020multilingual}.
Transformer-based methods have proved to be extremely useful in hate speech detection for low-resource languages.
\subsection{\textit{Classification Methods}}
In addition to  methods adopted for text feature extraction, abusive and hateful content detection methods also vary on the basis of the classification algorithms, i.e., Machine Learning and Deep Learning algorithms. 
In the related works, Machine Learning methods have been used mainly in combination with classical methods of feature extraction. 
Burnap \textit{et al.}~\citep{burnap2015cyber} uses ensembling of SVM, Logistic Regression (LR) and Random forest (RF) methods for hateful content detection on twitter. In this work, n-grams of hateful terms along with typed dependency gives best results for all the classification methods used for comparison. In Vigna \textit{et al.}~\citep{del2017hate} too n-grams of characters and words are used with SVM as classifier. They classify the content into three categories - No Hate, Weak Hate and Strong Hate. LSTM is taken as the second classifier for capturing long term dependencies in the content, but SVM outperforms LSTM in Vigna \textit{et al.}. Deep Learning methods have also been extensively explored for this task. These methods are either provided with the features extracted using classical methods or with the pretrained word or context embeddings from deep neural networks. In Mehdad \textit{et al.}~\citep{mehdad2016characters} RNN is used with character n-grams for hate speech detection, and RNN achieves better results than SVM in this work. CNNs are explored for hate detection in Gamback \textit{et al.}~\citep{gamback2017using} with Word2vec. Their method provides better results than LR classifier. Zhang \textit{et al.}~\citep{zhang2018detecting} uses mixture of CNN-GRU with word-embeddings to detect hate speech, which improves performance upto 10\% as compared to SVM baseline.
In light of the above discussion, it can be observed that selection of the classification method is very critical as none of the methods clearly outperform others. Though SVM generally performs better than other Machine Learning methods, the same distinction can not be made between Machine Learning and Deep Learning methods.
\subsection{\textit{Methods for Multilingual Low-resource Indic Languages}}
Abusive and hateful content has posed challenges for researchers in low-resource languages, such as Indic languages, due to code-mixing and multilingualism. Researchers have come up with different approaches to tackle this problem. Hande \textit{et al.}~\citep{hande2020kancmd} proposes Machine Learning methods for offensive language detection in Kannada language. They use TF-IDF based text features. Major work has been done recently to adapt Deep Learning techniques for multilingual Indic texts.  Chopra \textit{et al.}~\citep{chopra2020hindi} uses CNN-BiLSTM with author profiling and debiasing for hate detection in Hindi-English text. 

Numerous models, including Multilingual Bidirectional Encoder Representations from Transformers (M-Bert) \citep{devlin2019bert}, Multilingual Representations for Indian Languages (Muril) \citep{khanuja2021muril}, and XLM-R \citep{conneau2020unsupervised}  have become popular and have produced statistically significant improvements in abuse identification in Indic languages. Sharma \textit{et al.}~\citep{sharma2022ceasing} proposes Map Only Hindi method for code-switched text in the domain of hateful content detection by utilizing fine-tuned transformer-based models i.e., M-Bert and Muril. Due to their ability to detect abusive content on social media platforms, these methods have been used extensively for Indic languages in Sharif \textit{et al.}~\citep{sharif2021nlp}, Amjad \textit{et al.}~\citep{amjad2021threatening} and Velanker \textit{et al.}~\citep{velankar2023mono} too. They demonstrate that transformers outperform classical models in hate detection in Indic Languages.

\section{Methodology}
\label{sec:sa_meth}
\begin{figure*}
    \centering
    \includegraphics[width=16cm, height=9cm]{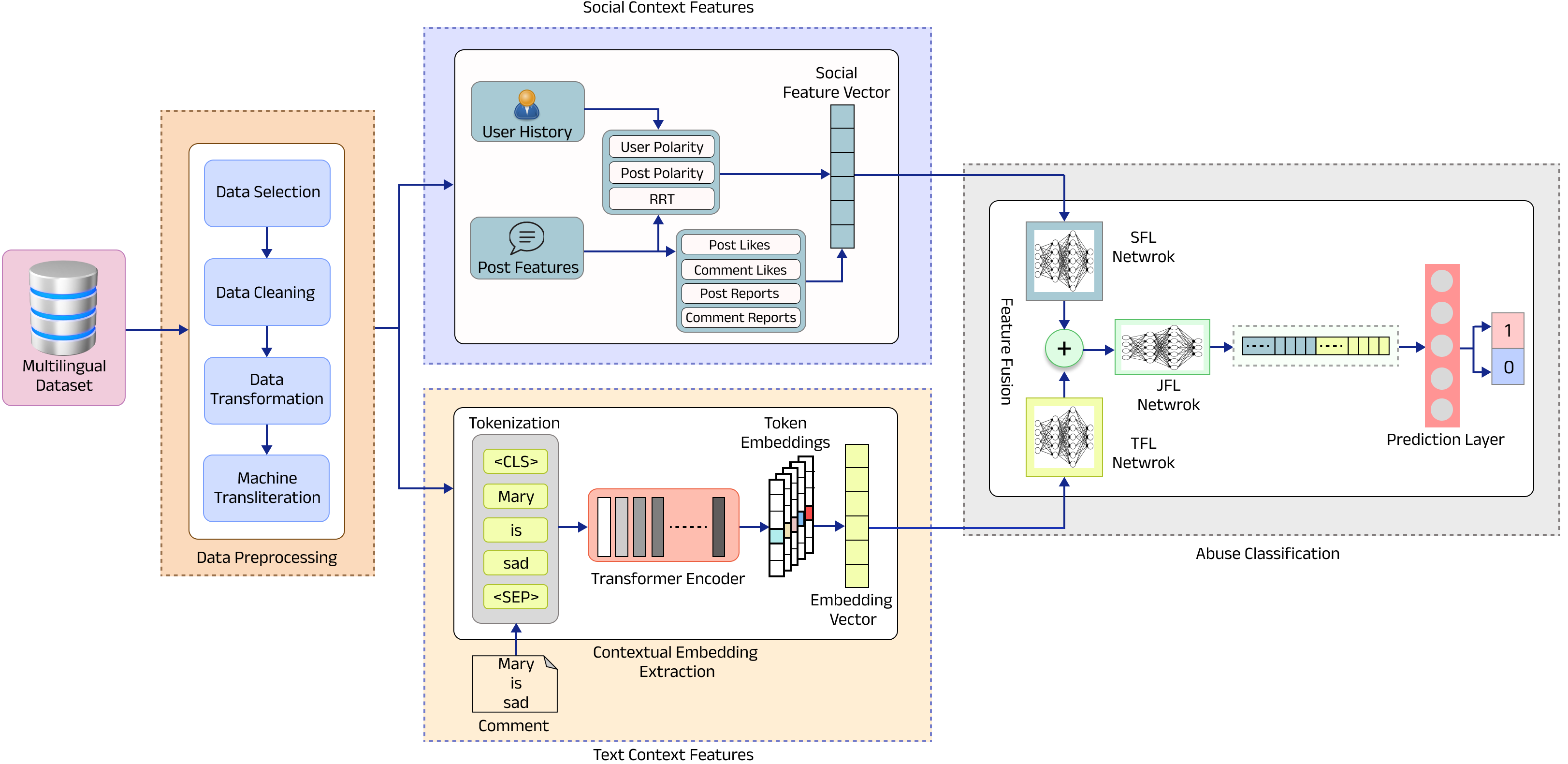}
    \caption{System Architecture}
    \label{fig:architecture}
\end{figure*}
A detailed discussion of our proposed methodology for abuse identification is presented in this section.
Figure-\ref{fig:architecture} demonstrates the system architecture of our proposed user-aware abusive content detection system. 
Proposed system comprises five major components: a) data preprocessing; b) social context features; c) text context features; d) abuse classification and e) ensemble learning.
In the data preprocessing module, the comment text set $C=\left\{c^i\right\}_{i=0}^K$ is cleaned and then transliterated to Roman script. Here $c^i$ denotes the $i$th comment in set $C$, $K$ is the cardinality of $C$. We denote the transliterated set of comments with $T_{c}=\left\{t_{c}^i\right\}_{i=0}^K$, where $t_{c}^i$ is $i$th transliterated comment. Transliteration is a salient feature of our proposed system. Comments from  set $T_c$ are used to get the contextual text embedding vector $\hat{v}$ for each comment $t_{c}^i$ $\in$ $T_c$ through transformer-based multilingual methods. We employ three transformer-based methods, Muril~\citep{khanuja2021muril}, M-Bert~\citep{devlin2019bert} and XLM-R~\citep{conneau2020unsupervised} for contextual embedding extraction. Parallelly, another module is trained with social context features. High-level representations of  text context features and social context features are then fused to get a joint representation $\hat{I}$. Architecture of social feature learning (SFL) module, text feature learning module (TFL) and joint feature learning (JFL) module is given in Figure-\ref{fig:classifier_module}. Finally, majority voting with confidence decision is employed for final classification. 

\subsection {\textit{Data Preprocessing}}
The data preprocessing step is pivotal as it sets the data up for future in-depth analysis in the most meaningful way possible. In layman's terms, uncleaned raw data is changed into cleaned data in this step. Before feeding the data to the model, data preprocessing steps, such as removing content with no text, removal of extra spaces and erroneous entries, are performed. Data preprocessing steps carried out in the proposed approach are elaborated as follows:

\subsubsection{\textit{Data Selection}} ShareChat IndoML Datathon NSFW (SCIDN)\footnote{Sharechat-indoml-datathon-nsfw-commentchallenge. URL: https://www.kaggle.com/competitions/
multilingualabusivecomment/data} dataset contains comments from more than 15 languages. Some of them are dialects of Hindi such as Haryanvi and Bhojpuri. We select comments from 12 prominent Indic languages, namely, Assamese, Bengali, English, Gujarati, Hindi, Kannada, Malayalam, Marathi, Odia, Punjabi, Tamil and Telugu. We take comments of 10 similar languages from the IIIT-D Multilingual Abusive Comment Identification dataset (MACI) \footnote{Iiit-d multilingual abusive comment identification. URL: https://www.kaggle.com/competitions/
iiitd-abuse-detection-challenge} except English and Punjabi as comments from these two languages are not present in that dataset. Table-\ref{table:desc_scidn},\ref{table:lang_desc_scidn},\ref{table:desc_maci} and \ref{table:lang_desc_maci} in Section-\ref{label:dataset_summary} give a detailed overview of both datasets.
\subsubsection{\textit{Data Cleaning}} In this step, we remove unexpected erroneous entries from the datasets such as rows with missing comments and missing feature values. Then comment-level cleaning is performed where we remove extra spaces, punctuation and digits from comments. Next, insignificant words are filtered out from comments utilizing our prepared set of such words.  
English words are taken from the NLTK library~\citep{bird2009natural}.
\subsubsection{\textit{Machine Transliteration}} \label{label:machine_trans}
Transliteration is a process to convert sentences written in one script to another script based on phonetic similarity. Many social media platforms give an option to write comments in native scripts as well as in Roman script.
In our experiments we observe that embeddings from transliterated comments achieve better results than the rest of the models.
In our pipeline, all comments are transliterated into Roman script. For a given comment c $\in$ C, $t_c$ denotes the comment after transliteration where $t_c \in T_c$.
\begin{equation}
t_c=IndicXlit (c)
\end{equation}
We use IndicXlit \citep{madhani2022aksharantar} method for transliteration. The largest publicly accessible parallel corpus with 26 million word pairings from Indic languages, Aksharantar, serves as the training data for IndicXlit. Dakshina \citep{roark2020processing} and Aksharantar benchmark are used for its evaluation. It employs six layers of encoders and decoders, 256-D input embedding vectors with four attention heads and a 1024 feedforward dimension for a total of 11M parameters. It currently supports 21 languages. We use transliterated comments for contextual embedding extraction for M-Bert and Muril, while for XLM-R we take comments in their original script as it shows better performance with them.
\subsubsection{\textit{Data Transformation}} Plenty of comments in the dataset contain emojis or emoticons. People use emojis to express anger, disgust, happiness and various other emotions while posting comments. Emojis can be utilized to get valuable emotional insights from a comment. We, therefore, transform all emojis into text. Next, uppercase characters are transformed into lowercase characters.

\subsection {\textit{Social Context Features}} \label{section:social_context_feature}

In numerous existing works related to abuse detection, only textual features have been extensively used with Machine Learning and Deep Learning models \citep{gamback2017using, park2017one}. A few incorporated user behavior as well \citep{chopra2020hindi}, \citep{pitsilis2018effective}. 
One objective of this work is to investigate the impact of both social context and textual features on abusive content detection methods.
Broadly, social context features include post features and user history. 
We present correlation scores of these features later in this section to establish their importance. Correlation scores show that our proposed feature, that is user-post polarity, has the strongest correlation with the class labels.

In the dataset, we have comments made on different posts, made by different users. Let’s assume a set of posts  $P=\left\{p^i\right\}_{i=0}^X$  and a set of users $U=\left\{u^j\right\}_{j=0}^Y$ , where $X$  represents the cardinality of set $P$ and $Y$ denotes the cardinality of set $U$.  
There may be more than one comment on a single post, and more than one comment made by a single user in the dataset. Hence, all the comments in the dataset can be represented either as the union of set of comments made on each post or the union of set of all comments made by each user as follows:
\begin{equation}
C=\left\{C_p^0\right\} \cup\left\{C_p^1\right\} \cup\left\{C_p^2\right\} \ldots \ldots . .\left\{C_p^X\right\}
\end{equation}
\begin{equation}
C=\left\{C_u^0\right\} \cup\left\{C_u^1\right\} \cup\left\{C_u^2\right\} \ldots \ldots . .\left\{C_u^Y\right\}
\end{equation}

where, $C$ denotes the set of all comments, $C_{p}^i$ denotes the set of comments made on a post $p^i$ and $C_{u}^j$ denotes the set of comments made by a user $u^j$.\\
We calculate two features, user polarity and post polarity.
For that we use the set of abusive words $A=\{a^i\}_{i=0}^{Q_1}$ collected by us, where $a$ denotes a single abusive word and $Q_1$ is the cardinality of set $A$.
We then prepare an extended set, $A_{ext}=\{a_{ext}^i\}_{i=0}^{Q_2}$, with different spellings of each abusive word from set $A$  by replacing a few letters with similar sounding letters, where $a_{ext}^i$ denotes a single abusive word and $Q_2$ is the cardinality of set $A_{ext}$.

A brief description of social context features is given in this section followed by their correlation with class labels.

\subsubsection{\textit{Post features}} 
Post features provide valuable information about the content, such as the number of likes and reports. Generally, the pattern of abusive content pertaining to the number of likes and reports would be different from non-abusive content. 
Therefore, we include post features in training. Following post features are available in both the datasets, namely, \textit{post id} ($p$), \textit{report count comment} ($r_{u}$), \textit{report count post} ($r_{p}$), \textit{like count comment} ($l_{c}$) and \textit{like count post} ($l_{p}$), where  $p\in P$. Apart from these, we determine two more post features, \textit{post polarity} and \textit{relative reporting tendency}, as follows:

\vspace{2mm}
\begin{enumerate}[label=\null]






    \item \textit{Relative reporting tendency (RRT):} 
    For any given comment $c$, let $r_{c}$ be the number of times it was reported, and, let $r_{p}$ be the number of times all the comments on that post were reported. Then  \textit{relative reporting tendency} ($r_{rt}$) can be determined as follows:
    \begin{equation}
    r_{rt}=\frac{r_c}{r_p}
    \end{equation}

    \item \textit{Post-polarity:} Post polarity gives the likelihood of a post to attract an abusive or non-abusive comment. For a given set of comments $C_{p}^i$  on a post $p^i$ , number of all abusive comments in this set can be written as $|\sigma_{L=1} C_{p}^i|$ and number of all non-abusive comments as $|\sigma_{L=0} C_{p}^i|$. Where $L = 1$ and $L = 0$  denote an abusive and non-abusive comment respectively and $\sigma$ denotes the selection operator.
     \textit{Post polarity} ($\phi_{p}$) can be calculated as:
     \begin{equation}
    \phi_{p}=\frac{\left|\sigma_{L=0}\left(C_p^i\right)\right|-\left|\sigma_{L=1}\left(C_p^i\right)\right|}{m}
    \end{equation}
    where, $m$ is the total of number comments on post $p^i$. Similar steps, as shown in Algorithm-\ref{algo:user-post_polarity} for user polarity feature can be repeated for determining post polarity.
\end{enumerate}

\subsubsection{\textit{User history}}
User history can be utilized to determine the likelihood of a comment to be abusive or non-abusive. Generally, a comment written by a user, who has made more abusive comments in the past, is more likely to be abusive. 

\vspace{2mm}
\begin{enumerate}[label=\null]
\item \textit{User polarity:} User polarity feature gives the user's tendency to make an abusive or non-abusive comment independent of the nature of the post. 
For a given set of comments $C_{u}^j$ made by a user $u^j$ , number of all abusive comments in this set can be written as $|\sigma_{L=1} C_{u}^j|$ and number of all non-abusive comments as $|\sigma_{L=0} C_{u}^j|$. 
\textit{User polarity} ($\phi_{u}$) can be calculated as:
  \begin{equation}
\phi_{u}=\frac{\left|\sigma_{L=0}\left(C_u^j\right)\right|-\left|\sigma_{L=1}\left(C_u^j\right)\right|}{m}
\end{equation}
where, $n$ is the total number of comments made by the user $u^i$. Algorithm-\ref{algo:user-post_polarity} gives the steps for calculation of $\phi_{u}$.
    
\end{enumerate}

\begin{algorithm}[h!]
    \caption{User-Post Polarity}
    \begin{tabular}{ll}
    \textit{Input:} & $T_{c,u}^j$:  set of comments made  by a user 
$u^j$\\
    & $T_{c,p}^i$  : set of comments on a post $p^i$\\
    &  $A_{ext}$  : set of abusive words \\
  \textit{Output:} & $\phi_u$: User Polarity of user u\\
  & $\phi_p$: Post Polarity of post p\\
  & $\phi_{up}$: Combined User and Post Polarity\\
  \textit{function:} & user\_post\_polarity($A_{ext},T_{c,u}^j, T_{c,p}^i$)
    \end{tabular}
    \label{algo:user-post_polarity}
    \begin{algorithmic}[1]
    \State{$\phi_u \gets $user\_polarity($A_{ext},T_{c,u}^j$)}
    \State{$\phi_p \gets $post\_polarity($A_{ext},T_{p,u}^i$)}
    \State{$\phi_{up} \gets \alpha (\phi_u)+(1 - \alpha )\phi_p$}
    \State{\Return$\phi_{up},\phi_u,\phi_p$}
    \end{algorithmic}
    \vspace{1mm}
    \begin{tabular}{ll}
    \textit{function:}& user\_polarity($A_{ext},T_{c,u}^j$)
    \end{tabular}
        \vspace{-2mm}
    \begin{algorithmic}[1]
    \State{$count_{abuse} \gets 0$}
    \State{$count_{non} \gets 0$}
    \State{\textbf{for  all $t_c \in T_{c,u}^j$ do}}
    \State{\hspace{3mm}flag $\gets$ 0}
    \State{\hspace{3mm}\textbf{for all $a_{ext}\in A_{ext}$ do}}
    \State{\hspace{5mm}\textbf{\textbf{if $a_{ext}$ in $t_c$ then}}}
    \State{\hspace{8mm}flag =1}
    \State{\hspace{8mm}\textbf{break}}
    \State{\hspace{5mm}\textbf{end if}}
    \State{\hspace{3mm}\textbf{end for}}
    \State{\hspace{3mm}\textbf{if flag is 1 then}}
    \State{\hspace{5mm}$count_{abuse} \gets count_{abuse}+1$}
    \State{\hspace{3mm}\textbf{else}}
    \State{\hspace{5mm}$count_{non} \gets count_{non}+1$}
    \State{\hspace{3mm}\textbf{end if}}
    \State{\textbf{end for}}
    \State{$
\phi_u^{\text {ext }} \leftarrow \frac{\text { count }_{\text {non }}-\text { count }_{\text {abuse }}}{\text { count }_{\text {non }}+\text { count }_{\text {abuse }}}
$}
\vspace{1mm}
\State{$
\phi_u \leftarrow max(\phi_u^{ext},\phi_u^{cls})$}
\State{\Return $\phi_u$}
    \end{algorithmic}
    \end{algorithm}
    
\subsubsection{\textit{Combined user-post polarity}} 
A user's behavior may depend on the type of post. For a post which receives mostly abusive comments, it is highly likely that any arbitrary user will also write an abusive comment on the post. However, it may not be true all the time as user tendency would also have a role to play to determine this. So, we calculate a combined user-post polarity feature which takes into account both polarities with appropriate weights given to each. User-post polarity is calculated as follows:
\begin{equation}
\phi_{up}=\alpha\left(\phi_{u}\right) + \left(1-\alpha\right)\phi_{p} 
\end{equation}

We test this feature with different weights given to each polarity for some random samples from the dataset. 
$\alpha$ is set to 0.47 as the model provides better results for this value of $\alpha$.

To calculate $\phi_u$ and $\phi_p$, we count the number of abusive and non-abusive comments for each user and post. Test data may not necessarily have comments from same users and posts as training data. So, we take the maximum of polarities determined using pre-classifier and lexicon matching approaches for 
test data. For this, we prepare a large set of  abusive words comprising words from all the 12 languages \citep{mathur2018did}, \citep{bird2009natural}. With the help of this set, we determine the values of $\phi_p^{ext}$ and $\phi_u^{ext}$. To determine polarities $\phi_p^{cls}$ and $\phi_u^{cls}$, Muril is taken as the pre-classfier. $\phi_u^{ext}$ and $\phi_p^{ext}$ denote the user and post polarities respectively, determined using extended abusive set. $\phi_u^{cls}$ and $\phi_p^{cls}$ denote the user and post polarities respectively, determined using the pre-classifier. 
For MACI dataset, we use only post-polarity feature as user-ids are not available in that dataset.

In user-post polarity function of Algorithm-\ref{algo:user-post_polarity}, in line 1 and line 2, user polarity and post polarity functions are called respectively which return the corresponding polarities. In line 3, combined user-post polarity is calculated as the weighted sum of individual polarities. In user polarity function of Algorithm-\ref{algo:user-post_polarity}, line 1,2 initialize the count of abusive and non-abusive comments to 0. In line 3-10, the set of all transliterated comments $T_{c,u}^j$, made by a user $u^j$ is iterated through to search the abusive words in each comment. Set of $A_{ext}$ is referred for this operation. In line 12-14, if an abusive word is found in the comment then the corresponding counter $count_{abuse}$ is incremented, otherwise the counter for non-abusive comments $count_{non}$ is incremented. In line 17, user polarity $\phi_u^{ext}$, which is determined by using $A_{ext}$, is calculated. Similarly, we calculate the count of abusive and non-abusive comments using the labels generated by the pre-classifier and these counts are used to calculate user polarity $\phi_u^{cls}$. In line 18, we take maximum of the two user polarities as the final user polarity $\phi_u$. Similarly, we calculate post polarity for each post.

\vspace{2mm}
\textit{Correlation: }
We use Point Biserial correlation \citep{kornbrot2014point} to represent the association of features with class labels. Point Biserial correlation coefficient is used to measure the correlation between a continuous and a dichotomous variable. It returns values which range from -1 to 1. A coefficient value of 1 represents a perfect positive correlation, -1 represents a perfect negative correlation, and 0 represents no correlation. Results are presented in Table-\ref{table:point_scidn} and Table-\ref{table:point_maci} for SCIDN and MACI datasets, respectively. User-post polarity shows the strongest correlation, whereas post-id has the weakest correlation with class labels in SCIDN. In MACI, post polarity has the strongest correlation, while post-id has the lowest correlation. Correlation for embeddings is calculated for Muril predicted labels.
We concatenate the selected features and create a social feature vector $\hat{s}$ . Next, the SFL module is trained on $\hat{s}$. We give a detailed description of the SFL module in Section-\ref{label:sfl}.

\begin{table}[h!]
\centering
\caption{Point Biserial correlation of features with class labels in SCIDN dataset}
\begin{tabular}{lc}
\toprule
\textbf{Feature}            & \textbf{Point Biserial} \\ \midrule
Post id                     & 0.002                   \\ 
Like count comment          & -0.097                  \\ 
Like count post             & 0.136                   \\ 
Report count comment        & -0.010                  \\ 
Report count post           & 0.105                   \\ 
Relative reporting tendency & -0.019                  \\ 
Post polarity               & -0.906                  \\ 
User id                     & 0.157                   \\ 
User polarity               & -0.752                  \\ 
User-post polarity          & -0.911                 \\ 
Contextual Embeddings       & 0.873                   \\ \bottomrule
\end{tabular}
\label{table:point_scidn}
\end{table}

\begin{table}[h!]
\centering
\caption{Point Biserial correlation of features with class labels in MACI dataset}
\begin{tabular}{lc}
\toprule
\textbf{Feature}            & \textbf{Point Biserial} \\ \midrule
Post id                     & -0.007                  \\
Like count comment          & -0.080                  \\ 
Like count post             & -0.055                  \\ 
Report count comment        & -0.022                  \\ 
Report count post           & -0.007                  \\ 
Relative reporting tendency & -0.028                  \\ 
Post polarity               & -0.857                 \\ 
Contextual Embeddings       &  0.752                   \\ \bottomrule
\end{tabular}
\label{table:point_maci}
\end{table}

\subsection {\textit{Text Context Features}} \label{section:text_context_feature}
In this section, we present a detailed overview of the text context features. It comprises two subsections: a) data augmentation, and  c) contextual embedding extraction. Data augmentation is employed by inserting additional synthetic comments in the dataset using a random sample of original comments from the training data. It is done with the objective of dampening the effect of misspelled abusive words. Next, contextual embeddings are extracted and fed to the TFL module. The rest of this section describes these steps in detail.


\subsubsection{\textit{\hspace{0.3mm}Data Augmentation}} 
We observe that the same words were transliterated differently in different comments. Users, too, tend to make spelling mistakes while writing comments. 
To handle this problem, we use the data augmentation technique. 
For that, we use the extended set of abusive words $A_{ext}$. Set $T_c$ is divided into train set $T_c^{train}$ and test set $T_c^{test}$ as shown in Equation 8. 
\begin{equation}
T_c=T_c^{train } \cup T_c^{test }
\end{equation}
From the train set $T_c^{train}$, we sample a random set of non-abusive comments having same cardinality as $A_{ext}$. Each abusive word $a_{ext} \in A_{ext}$ is inserted into a comment. Original comments are also retained in train set $T_c^{train}$. Since insertion of the abusive word makes the comment abusive, label is also inverted for these synthetic comments.
This process is repeated for each language and we get an augmented set, $T_c^ {aug}$, which is combined with train set to get the final augmented training set $T_c^{train\_aug}$ as shown in Equation 9. 

\begin{equation}
T_c^{train\_aug}=T_c^ {train} \cup T_c^ {aug}
\end{equation}
This set of additional comments, $T_c^{aug}$, helps the model in two aspects, first, training samples have increased for each language while covering spelling errors in abusive words, secondly, the model can give more attention to words which make a comment abusive since the training data contains the non-abusive copy of the same comments as well. We argue that synthetic comments still make sense semantically because an abusive word is added at the starting of each sentence. Generally, most of the abusive words are person oriented, hence, adding them at the starting does not alter the grammar of the sentences, and semantics of the sentences are retained while the polarity changes from non-abusive to abusive. Steps of data augmentation are presented in Algorithm-\ref{algo:data_aug}.

\begin{algorithm}[!h]
    \caption{Data Augmentation}
    \begin{tabular}{ll}
    \textit{Input:} & $T_{c}^{train}$:  training set of comments
$u_j$\\
    &  $A_{ext}$  : extended set of abusive words \\
  \textit{Output:} 
  & $T_c^{train\_aug}$: augmented training set of comments\\
  \textit{function:} & augment($T_{c}^{train},A_{ext}$)
    \end{tabular}
    \label{algo:data_aug}
    \begin{algorithmic}[1]
    \State{$ 
T_c^{aug} \gets $\{ \}}
\State{\textbf{for  
$i \in language$ do}
}
\State{\hspace{3mm}\textbf{for each 
$a_{ext} \in A_{ext}^i$ and $t_c \in T_c^{train,i}$   do}
}
\State{\hspace{5mm}$ 
T_c^{aug} \gets T_c^{aug} \cup t_c.insert(a_{ext})$
}
\State{\hspace{3mm}\textbf{end for}}
\State{\textbf{end for}}

\State{$T_c^{train\_aug} \gets T_c^{train} \cup T_c^{aug}$
}
    \State{\Return $T_c^{train\_aug}$}
    \end{algorithmic}
    \end{algorithm}
 
In Algorithm-\ref{algo:data_aug}, line 1 initializes $T_c^{aug}$ as an empty set. In line 2-6, for each language, we iterate over a fixed number of comments equal to the number of abusive words for that language. Then, we insert an abusive word in the comment and the comment is added to $T_c^{aug}$ set. In line 7, we add the augmented set $T_c^{aug}$ to the training set. Finally, the augmented training set $T_c^{train\_aug}$ is returned.

\subsubsection{\textit{Contextual Embedding Extraction}}\label{section:contextual_embedding}
We employ three different methods to generate embeddings for a tokenized sequence of words, namely, Muril, XLM-R and M-Bert.
The general workflow for contextual embedding extraction, irrespective of the method we use, can be summarized using the example of the case involving Muril and transliterated comments.
First, tokenization of a transliterated comment $t_c \in T_c$ is done. A tokenized comment is represented as a sequence of tokens $W=\{w^i\}_{i=0}^{len}$, where $W$ denotes the set of tokens, $w$ is a single token and $len$ denotes the total number of tokens in the comment.
Next, two unique tokens, class [CLS] token and separator [SEP] token, are inserted at the beginning and at the end, respectively. 
The set of tokens is then converted into a set of integers represented by $word\_ids$ in Equation 10. 
\begin{equation}
words\_ids=tokens\_to\_ids (W)
\end{equation}
Subsequently, the sequence length of the input is fixed. We perform experiments for input sequence lengths of 128 and 64, in the ensemble framework. Any comment with number of tokens exceeding the sequence length is truncated. For comments shorter than the sequence length, padding tokens are added. Then, we create $input\_mask$ by adding 1s a number of times which is equal to the length of the tokenized sequence, followed by padding by 0s to the appropriate length, if required. This helps the model to distinguish between original tokens and padding tokens. Then, we create $input\_type\_ids$ which helps in determining the positional encoding of a sentence. Finally, we pass all three inputs to the encoder as shown in Equation 11, which generates the sentence embedding $E \in \mathbb{R}^D$ and the last hidden state matrix $H \in \mathbb{R}^{ l \times   D}$.
\begin{equation}
E, H \leftarrow encoder(words\_ids, input\_mask, input\_type\_ids) 
\end{equation}
Here, $l$ is the sequence length and $D$ is the dimension of each token embedding which  is 768, $\mathbb{R}$ symbolizes the set of real numbers. We use last hidden state to create the text embedding vector which is discussed in detail in Section-\ref{label:tfl}. We use encoders of the three aforementioned methods to generate the embeddings. Only base versions of these methods are used, which employ 12 encoder layers, with 12 self-attention heads in each layer.

\subsection {\textit{Abuse Classification}} \label{section:classifier}
\begin{figure*}
    \centering
    \includegraphics[width=15cm, height=8.3cm]{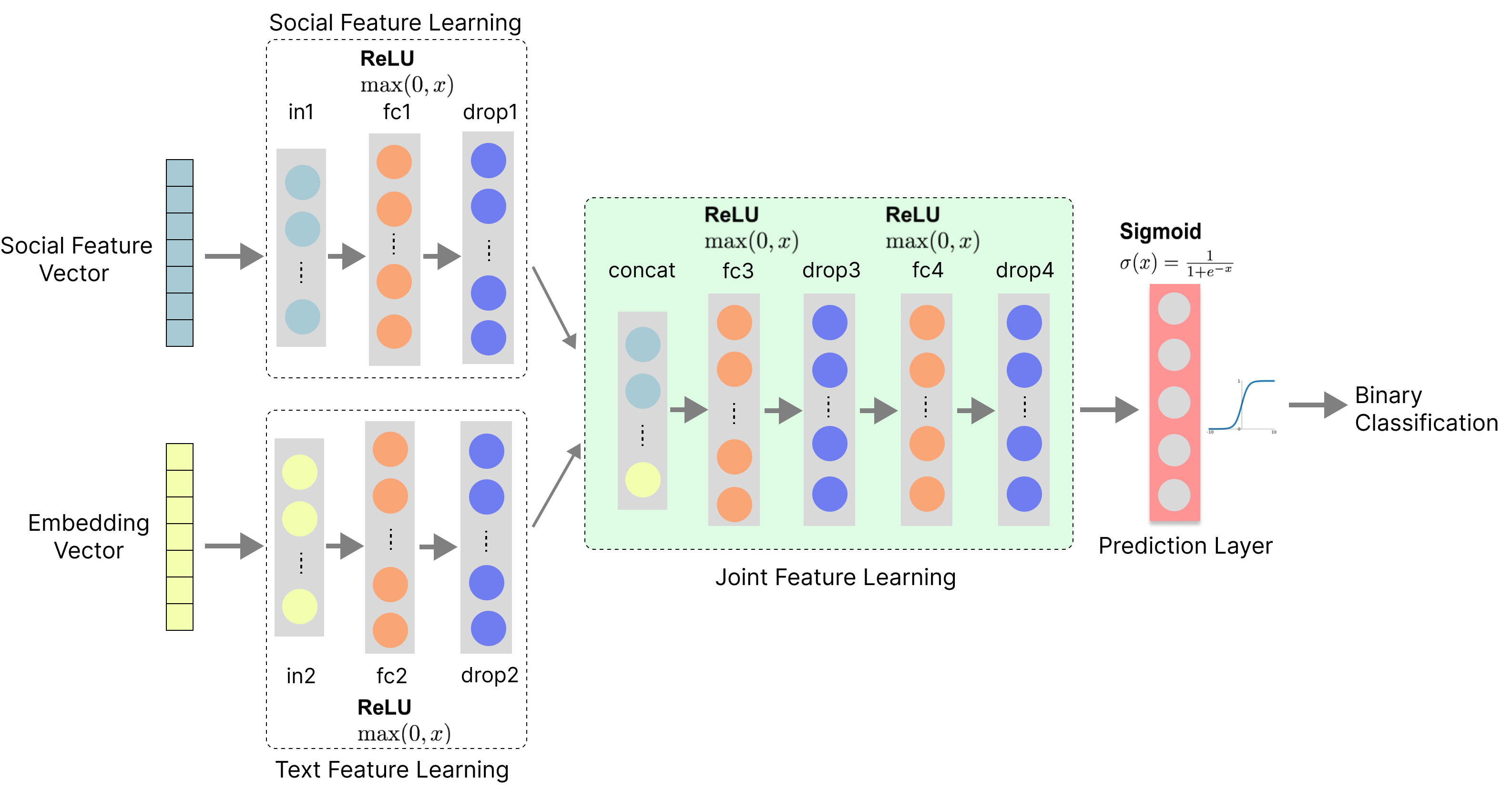}
    \caption{Architecture of abuse classification module}
    \label{fig:classifier_module}
\end{figure*}

In this section, we discuss the abuse classification module. Various components of the abuse classification module are shown in Figure-\ref{fig:classifier_module}. Four components constitute this module, namely, a) social feature learning; b) text feature learning; c) joint feature learning and d) prediction layer.

\vspace{2mm}
\subsubsection{
\textit{Social Feature Learning (SFL)}} \label{label:sfl}
We use six social context features, namely, report count post ($r_p$), like count comment ($l_c$), like count post ($l_p$), relative reporting tendency ($r_{rt}$) and combined user-post polarity ($\phi_{up}$). Here $r_{rt}$ and $\phi_{up}$ are the manufactured features. For MACI dataset we use report count comment ($r_c$) instead of $r_p$, as its correlation coefficient is higher as shown in Table-\ref{table:point_maci}.
We first scale these features using min-max normalization. Min-max normalization method is represented in Equation 12 .
Let  $x$, $x_{min}$ and $x_{max}$ denote the original value of a feature instance, minimum value for that feature, and maximum value for that feature, respectively. Then, min-max normalized value $x_{norm}$ of $x$ would be:
\begin{equation}
x_{norm }=\frac{x-x_{min }}{x_{max }-x_{min }}
\end{equation}
Next, a joint representation of social context features is created by concatenating these features as shown in Equation 13. We call this joint representation as social feature vector, denoted by $\hat{s}$.
\begin{equation}
\hat{s}=f_{cat}\left(r_p \oplus l_c \oplus l_p \oplus r_{rt} \oplus \phi_{up}\right)
\end{equation}
Here, $\hat{s} \in \mathbb{R}^M$, $M$ denotes the dimension of $\hat{s}$ vector, $f_{cat}$ denotes the concatenation function and $\oplus$ denotes the concatenation operation.
Vector $\hat{s}$ is then passed to input layer $in1$ of the SFL module which instantiates the tensor $\overline{\mathrm{T}}$($\hat{s}$). We pass the tensor $\overline{\mathrm{T}}$($\hat{s}$) to a fully connected dense layer $fc1$ to obtain the high-level representation, $\hat{S}$, of the social feature vector $\hat{s}$ as given in Equation 14.
\begin{equation}
\hat{S}=fc1\left(units = D_1\right)(\overline{\mathrm{T}}(\hat{s}))
\end{equation}
The layer $fc1$ performs dot product between $\overline{\mathrm{T}}$($\hat{s}$)  and weight matrix $W_1^{(r)}$. The operations performed by $fc1$ are represented by Equation 15.
\begin{equation}
fc1: \quad \hat{S}=\sigma_{relu }\left(W_1^{(r)} \cdot \overline{\mathrm{T}}(\hat{s})+b_1\right)
\end{equation}
Here, $\hat{S} \in \mathbb{R}^{D_1}$, $D_1$ denotes the dimension of $\hat{S}$ which is 16, $W_1^{(r)}$ is the weight matrix of dimension $r$ which is $D_1\times M$ i.e., 16×5, $b_1$ is bias vector of size $D_1$ and $\sigma_{relu}$ denotes the activation function.
We use Rectified Linear Unit (ReLU) as the activation function for $fc1$ . It adds non-linearity in the model. The ReLU function is represented in Equation 16 with $x$ as input.
\begin{equation}
\sigma_{relu }(x)= \begin{cases}0, & if\ \ x<0 \\ x, &   if\ \ x \geq 0\end{cases}
\end{equation}
ReLU makes the computation of gradient extremely simple as it gives either 0 or $x$ as output which depends on the sign of $x$. 0 is returned for non-positive values, otherwise $x$ is returned. It helps in speeding up the training of the neural network. To prevent overfitting, 20\% of the neurons are randomly dropped. 
Next, $\hat{S}$ is fused with the high-level representation of the text embedding vector $\hat{v}$ for joint learning. Joint feature learning is explained in detail in Section-\ref{label:jfl}.

\subsubsection{\textit{Text Feature Learning 
 (TFL)}} \label{label:tfl}
This section illustrates the text feature learning module. 
As mentioned in Section-\ref{section:contextual_embedding}, transformer-based multilingual methods are used for contextual embedding extraction. They generate the sentence embedding $E\in\mathbb{R}^D$ and the last hidden state matrix $H \in \mathbb{R}^{l\times D}$. We use last hidden state $H \in \mathbb{R}^{l \times D}$ for our experiments. As shown in Equation 17, all the token embeddings from the last hidden matrix $H$ are stacked horizontally to create embedding vector $\hat{v}$.
\begin{equation}
\hat{v}=Reshape(H)
\end{equation}
Here, $\hat{v} \in \mathbb{R}^N$, where $N$ is the dimension of text embedding vector. N, dimension of vector $\hat{v}$, depends on the sequence length. We illustrate the TFL module using a sequence length of 128. Thus $N$ becomes 98,304.
Next, embedding vector $\hat{v}$ is passed to the input layer $in2$ of the TFL module which instantiates the tensor $\overline{\mathrm{T}}$($\hat{v}$). We pass the tensor $\overline{\mathrm{T}}$($\hat{v}$) to the fully connected dense layer $fc2$ to obtain the high-level representation, $\hat{V}$, of the embedding  vector $\hat{v}$ as given in Equation 18.
\begin{equation}
\hat{V}=fc2\left( units =D_2\right)(\overline{\mathrm{T}}(\hat{v}))
\end{equation}
The layer $fc2$ performs dot product between $\overline{\mathrm{T}}$($\hat{v})$  and weight matrix $W_2^{(y)}$. All the operations performed by $fc2$ are represented by Equation 19.
\begin{equation}
fc2: \quad \hat{V}=\sigma_{r e l u}\left(W_2^{(y)} \cdot \overline{\mathrm{T}}(\hat{v})+b_2\right)
\end{equation}
Here, $\hat{V} \in \mathbb{R}^{D_2}$, 
$D_2$ denotes the dimension of $\hat{V}$ which is 768, $W_2^{(y)}$ is the weight matrix of dimension $y$ which is $D_2 \times N$, i.e., 768×98,304, $b_2$ is bias vector of size $D_2$ and $\sigma_{relu}$ denotes the activation function. ReLU is used in $fc2$ too as activation function. As in the SFL module, we randomly drop 20\% of the neurons in TFL.
\vspace{2mm}
\subsubsection{\textit{Joint Feature Learning (JFL)}} \label{label:jfl}
 This module employs a feature fusion mechanism to obtain a joint representation of the learnt social feature vector $\hat{S}$ and the text embedding vector $\hat{V}$. Then, module learns this joint representation. As shown in Equation 20, we first pass $\hat{V}$ and $\hat{S}$ to $concat$ layer which concatenates both vectors and outputs a joint feature vector $\hat{I}$.
\begin{equation}
\hat{I}=concat(\hat{V} \oplus \hat{S})
\end{equation}
Here, $\hat{I} \in \mathbb{R}^{D_3}$, $D_3$ is the dimension of the joint feature vector which is 784, $concat$ denotes the concatenation layer and $\oplus{}$ denotes the concatenation operation. This 784-$D$ vector is passed to two fully connected layers sequentially, with a dropout layer after each fully connected layer. Equation 21, 22, 23 and 24 represent further operations on the joint vector.
\begin{equation}
\hat{i}_1=f c3\left( units =D_4\right)(\hat{I})
\end{equation}
\begin{equation}
fc3: \quad i_1=\sigma_ {relu }\left(W_3^{\left(z_1\right)} \cdot \hat{I}+b_3\right)
\end{equation}
\begin{equation}
\hat{i}_2=f c 4\left( units =D_4\right)\left(\hat{i}_1\right)
\end{equation}
\begin{equation}
fc4: \hat{i}_2=\sigma_ {relu }\left(W_4^{\left(z_2\right)} \cdot \hat{i}_1+b_4\right)
\end{equation}
We obtain two intermediate vector outputs  of dimension $D_4$, $\hat{i_1}$ and $\hat{i_2}$,   from $fc3$ and $fc4$ layers respectively. Based on our experiments, we set $D_3$ as 100. $W_3^{\left(z_1\right)}$ and $W_4^{\left(z_2\right)}$ are weight matrices for $fc3$ and $fc4$ of dimension $z_1$ and $z_2$ which are 100×784 and 100×100 respectively. $b_3$ and $b_4$ biases of 100-$D$ are used in $fc3$ and $fc4$. 
Next, we pass intermediate vector $\hat{i_2}$ to the prediction layer for final classification. 

\vspace{2mm}
\subsubsection{\textit{Prediction Layer}}
We generate class labels for each comment in multilingual comment set $C$.
To obtain the predicted class label, we pass the output of the JFL module, i.e., intermediate vector $\hat{i_2}$, to the prediction layer as represented in Equation 25. 
\begin{equation}
Y_{pred }=\textit {Dense}(\textit {units}=1)\left(\hat{i}_2\right)
\end{equation}
Here, $Y_{pred}$ is the predicted probability, and $Dense$ denotes the prediction layer. Next, we apply logistic sigmoid activation on the prediction layer to get a predicted probability value between 0 and 1. This type of sigmoid activation function can be described by the mathematical formula given in Equation 26.
\begin{equation}
\sigma_{\text {sig }}(x)=\frac{1}{1+e^{-x}}
\end{equation}
This sigmoid function accepts any real-valued input and returns a real-value between 0 and 1. 
Next, loss is calculated to check the difference between the predicted probability $Y_{pred}$ and the ground truth $L_{gt}$ . As we have just two classes, we take binary cross entropy loss $L_{BCE}$  as the loss function for our model. The binary cross entropy loss function computes loss by taking the following average:
\small
\begin{equation}
L_{BCE}=\frac{1}{B} \sum_{i=0}^B \left(L_{gt}^i-1\right) \log \left(1-Y_{pred }^i\right)-L_{gt}^i \log \left(Y_{pred}^i\right)
\end{equation}
\normalsize
where, $B$ denotes the batch size. We use the Adam optimizer to obtain the gradients of loss function and weights are updated accordingly.
Then, probability is converted to one of the binary labels 0 and 1 using $set\_label$ function as represented in Equation 28.
\begin{equation}
 Label = set\_label \left(Y_{pred} \geq\right.  threshold, 1, 0) 
\end{equation}
If probabilities are greater than or equal to threshold, they are given the label 1, otherwise the label 0 is assigned. Label 1 represents the abusive class and label 0 represents the non-abusive class.

\subsection {\textit{Ensemble Learning}}
   
Ensemble learning is used to achieve better predictive performance by capitalizing the predictive effectiveness of different classification algorithms. It can be considered as a technique to create a strong classifier by combining multiple base classifiers.
In this section, we discuss ensemble learning in detail. This section consists of three subsections: a) methods used in ensembling;  b) majority voting and c) confidence decision. Algorithm-\ref{algo:majority_voting} presents the sequence of steps performed in ensemble learning.

\subsubsection{\textit{Methods}}
We employ three methods for embedding extraction for ensemble learning. We take two text embedding vectors for two different sequence lengths from each of the three different transformer-based methods as depicted by Equation 29.
\begin{equation}
\hat{v}=Reshape\left({TF}\left(t_c\right)_{128,64}\right)
\end{equation}
Here, $TF = \{Muril, MBert, XLMR\}$ represents all the contextual embedding generation methods used to transform the comments to embeddings.
These two embedding vectors are generated for 64 and 128 input sequence lengths. Predictions are taken for all the text embedding vectors, thus giving us a total of six abuse classification methods, one for each embedding. Six results, therefore, are taken for a single comment $t_c \in T_c$.

\begin{algorithm}[h!]
    \caption{Majority voting with confidence decision}
    \begin{tabular}{ll}
    \textit{Input:} & $L^i$:  set of predicted labels for 
ith comment \\
    & $Y_{p}^i$  : set of predicted probabilities of classes\\
  \textit{Output:} &  
$F_{label}^i$ :  final label for $i$th comment\\ 
\\
  \textit{function:} &     majority\_voting ($L^i,Y_p^i$)
    \end{tabular}
    \label{algo:majority_voting}
    \begin{algorithmic}[1]
    \State{\textbf{if  
$\Sigma_{j=0}^G Label_1^j > 3$ then}}
    \State{\hspace{3mm}$F_{label}^i \gets 1$}
    \State{\textbf{else if 
 $\Sigma_{j=0}^G Label_1^j \le 2$ then}}
    \State{\hspace{3mm}$F_{label}^i \gets 0$}
    \State{\textbf{else}}
    \State{\hspace{3mm}$F_{label}^i \gets confidence\_decision(Y_p^i)$}
    \State{\textbf{end if}}
    \State{\Return $F_{label}^i$}
    \end{algorithmic}
    \vspace{1mm}
    \begin{tabular}{ll}
    \textit{function:}& confidence\_decision($Y_p^i$)
    \end{tabular}
        \vspace{-2mm}
    \begin{algorithmic}[1]
    \State{\textbf{if   $\Sigma_{j=0}^3 dist(Y_{pred_1}^j)> \Sigma_{j=0}^3 dist(Y_{pred_0}^j)$then}}
    \State{\hspace{3mm}\textbf{return} 1}
    \State{\textbf{else if  
 $\Sigma_{j=0}^3 dist(Y_{pred_1}^j)< \Sigma_{j=0}^3 dist(Y_{pred_0}^j)$then}}
     \State{\hspace{3mm}\textbf{return} 0}
    \State{\textbf{else}}
         \State{\hspace{3mm}\textbf{return} best\_model($Y_p^i$ )}
    \State{\textbf{end if}}
    \end{algorithmic}
    \end{algorithm}

\subsubsection{\textit{Majority Voting}}
We use majority voting for final classification. If majority of the models agree on a label, that label is taken as the final label. Majority voting is mainly used when models’ outputs are independent. As shown in Equation 30, we pass the set of predicted labels $L^i=\{Label^j\}_{j=0}^G$   and predicted probabilities set for the $i$th comment to the $majority\_voting$ function to obtain the final label.
\begin{equation}
F_{label}^i= majority\_voting \left(L^i, Y_p^i\right)
\end{equation}
Here, $F_{label}^i$ is the final label for comment $c^i$, $L^i$ denotes the set of predicted labels for a single comment $c^i$, $Label^j$ denotes the $j$th label from the set $L^i$, and  $G$ is the cardinality of the set which is 6. $Y_p^i$ is the set of predicted probabilities for the $i$th comment.
Since we have an even number of models, there may be a tie among the models. For this situation, we employ confidence decision which is described next.

\subsubsection{\textit{Confidence Decision}}
In case of an even number of models in ensemble learning, a situation of tie occurs in binary classification when half of the models predict one class, and the other half predict the other class. We employ the confidence decision technique based on sum of distance of probabilities from threshold. If for a label, this sum is higher than that of the other label, this label is selected as the final label.
\begin{equation}
F_{label }^i= confidence\_decision \left(Y_p^i\right)
\end{equation}
The function of the $confidence\_decision$ can be written as:
\small
\begin{equation}
f(Y^i_p)= \begin{cases}1, & { if }\ \
\Sigma_{j=0}^3 \operatorname{dist}\left(Y_{pred_1}^j\right)>\Sigma_{j=0}^3 \operatorname{dist}\left(Y_{pred_0}^j\right) \\ 0, &  {otherwise}  \end{cases}
\end{equation}
\normalsize
where, $Y_p^i=\{Y_{pred}^j\}_{j=0}^G$ denotes the resultant set of probabilities for comment $c^i$ for 6 models, G is the cardinality of the set which is 6. $Y_{pred_1}$ is the probability for 1 to be the label, and $Y_{pred_0}$ is the probability for 0 to be the label. Distance function $dist$ calculates the distance of a probability from the threshold value.
In case of tie in sum, which is rare, we take the label given by Muril embedding model of 128 sequence length as the final label as it is the best performing individual model.

In majority voting function of Algorithm-\ref{algo:majority_voting}, Line 1 and Line 3 compare and calculate the number of 1s from the set of predicted labels for a single comment. Line 6 calls the confidence decision function in case of a tie between the number of labels of the two classes. In the confidence decision function of Algorithm-\ref{algo:majority_voting}, line 1 and line 3 calculate and compare sum of distance of probabilities from threshold for two labels. Based on these comparisons, a label is returned, otherwise line 6 returns the label given by the best model.

\section{Experimental Evaluations}
This section first discusses the experimental setup and then shows the experimental results to demonstrate the effectiveness of our technique.

\subsection {\textit{Experimental Setup}}
In this section, we first describe datasets used for experimental evaluations. Next, different methods used for comparative analysis are discussed. We also discuss the evaluation metrics used for evaluating the performance of different methods.

\subsubsection{\textit{Summarization of Datasets}}\label{label:dataset_summary}
We perform abusive language detection on two different datasets, namely, SCIDN and MACI. SCIDN is a publicly available multilingual dataset from ShareChat. MACI is a similar multilingual text-based dataset taken from Moj.
\begin{enumerate}[ label=\alph*) ]
\item{
        \textit{SCIDN}\footnote[4]{Sharechat-indoml-datathon-nsfw-commentchallenge. URL: https://www.kaggle.com/competitions/
multilingualabusivecomment/data}: ShareChat IndoML Datathon NSFW (SCIDN) dataset has comments in more than 15 Indic languages, which have been collected from the ShareChat social media application. 
        The SCIDN dataset has more than 1.5M human annotated comments. 
        We first take comments for 12 different Indic languages, Assamese (as), Bengali (bn), English (en), Gujarati (gu), Hindi (hi), Kannada (kn), Malayalam (ml), Marathi (mr), Odia (or), Punjabi (pn), Tamil (ta) and Telugu (te). Next, we remove rows with erroneous and null comments. Table-\ref{table:desc_scidn} and \ref{table:lang_desc_scidn} present a detailed overview of the SCIDN dataset.
         
        \begin{table}[h!]
        \centering
        \caption{SCIDN dataset description}
        \begin{tabular}{ll}
        \toprule
        \textbf{Label} & \textbf{Percentage} \\ \midrule
        Non-Abusive    & 333,902 (50.36\%)    \\ 
        Abusive        & 329,157 (49.64\%)    \\ \bottomrule
        \end{tabular}
        \label{table:desc_scidn}
        \end{table}
        
        \begin{table}[h!]
        \centering
        \caption{Language-wise description of SCIDN dataset}
        \begin{tabular}{ll|ll}
        \toprule
        \textbf{Language} & \textbf{Samples} & \textbf{Language} & \textbf{Samples} \\ \midrule
        Hindi             & 266,537           & Kannada           & 29,679            \\ 
        English           & 79,618            & Bengali           & 22,237            \\ 
        Telugu            & 61,292            & Odia              & 17,355            \\ 
        Punjabi           & 54,964            & Marathi           & 16,177            \\ 
        Malayalam         & 52,790            & Gujarati          & 10,969            \\ 
        Tamil             & 49,989            & Assamese          & 1,453             \\ \bottomrule
        \end{tabular}
        \label{table:lang_desc_scidn}
        \end{table}
        
        }

        \item{
        \textit{MACI}\footnote[5]{Iiit-d multilingual abusive comment identification. URL: https://www.kaggle.com/competitions/
iiitd-abuse-detection-challenge}: IIIT-D Multilingual Abusive Comment Identification dataset was taken from the IEEE BigMM challenge which aims to improve abusive content detection on social media for low-resource Indic languages. Comments are collected from the Moj application. MACI contains comments from more than 10 Indic languages. The dataset is fairly balanced for instances of the abusive and non-abusive class. The raw dataset comprises 665K human annotated comments labeled as abusive or non-abusive. In Table-\ref{table:desc_maci} and \ref{table:lang_desc_maci}, we give a detailed description of the MACI dataset.

        \begin{table}[h!]
        \centering
        \caption{MACI dataset description}
        \begin{tabular}{ll}
        \toprule
        \textbf{Label} & \textbf{Percentage} \\ \midrule
        Non-Abusive    & 352,376 (52.98\%)    \\ 
        Abusive        & 312,656 (47.02\%)    \\ \bottomrule
        \end{tabular}
        \label{table:desc_maci}
        \end{table}
        
        \begin{table}[h!]
        \centering
        \caption{Language-wise description of MACI dataset}
        \begin{tabular}{ll|ll}
        \toprule
        \textbf{Language} & \textbf{Samples} & \textbf{Language} & \textbf{Samples} \\ \midrule
        Hindi             & 307,179           & Telugu            & 97,011            \\ 
        Marathi           & 72,044            & Bengali           & 22,835            \\ 
        Tamil             & 69,496            & Odia              & 10,973            \\ 
        Kannada           & 13,943            & Malayalam         & 40,959            \\ 
        Gujarati          & 8,828             & Assamese          & 2,780             \\ \bottomrule
        \end{tabular}
        \label{table:lang_desc_maci}
        \end{table}
        }
    \end{enumerate}
    
\subsubsection{\textit{Comparison Methods}}
    We perform comparative analysis to evaluate the performance of our model. For this purpose, we compare the results of our model with the following methods:
\begin{enumerate}[ label=\alph*) ]
\item{\textit{Classical Methods:}
For comparison with classical methods, we first take textual features from two classical methods, count-based and frequency-based surface level feature methods. These features are then used with SVM, LR and Random-Forest Machine Learning models. We present results only for the best performing method for a given classical feature set. SVM is found to give better results with character-level features for both feature extraction techniques. Next, we take word-embeddings based method for text feature extraction. Unlike count-based and frequency-based methods which generate large sparse vectors, word-embeddings are dense vectors of fixed size. For comparison, we take non-contextual word-embeddings which are specifically trained on our datasets. These embeddings are fed to MLP and LSTM Deep Learning models for predictions. MLP is found to outperform LSTM. The best performing method's result is shown for word-embeddings too. Results are shown in Table-\ref{table:compare_scidn}  and Table-\ref{table:compare_maci} in Section-\ref{label:experiment_results}.
}

\item {\textit{State-of-the-art Baseline Methods:}
We evaluate the performance of our method against recent transformer-based multilingual baseline methods. Muril, XLM-R and M-Bert embeddings are taken and used for classification. Muril is found to give better results for SCIDN and XLM-R performs the best on MACI. Comparison results are shown in Table-\ref{table:compare_scidn}  and Table-\ref{table:compare_maci} in Section-\ref{label:experiment_results}.} 
 Next we compare the efficacy of the proposed method with existing research works. Results of our method are compared with Sazzed \textit{et al.} \citep{sazzed2021abusive}, Modha \textit{et al.} \citep{modha2020detecting} and Bansal \textit{et al.} \citep{bansal2022transformer} methods. Detailed results are shown in Table-\ref{table:compare_scidn}  and Table-\ref{table:compare_maci}.
 Lastly, we perform a language-wise comparative analysis between  transformer-based methods and the proposed method. The proposed method outperforms existing methods for every language as shown in Table-\ref{table:scidn_lang}  and Table-\ref{table:maci_lang}.
\end{enumerate}

\subsubsection{\textit{Evaluation Metrics}}
Abusive content detection is mainly modeled as a classification problem. Comments are classified into two different classes, abusive and non-abusive. We aim to show that our method gives results with the same effectiveness across multiple evaluation metrics. Hence, we use four evaluation metrics for comparison, namely, Accuracy (Acc), Precision (P), Recall (R) and F1-score (F1). F1-score can be a good measure even in case of unbalanced classes.

Let us suppose that $L_{TP}$, $L_{FP}$, $L_{TN}$, and $L_{FN}$ are the sets of labels predicted correctly as abusive, predicted falsely as abusive, predicted correctly as non-abusive and predicted incorrectly as non-abusive, respectively. Precision is calculated by dividing the number of elements in the set of correctly predicted abusive labels $L_{TP}$  by the total number of actual abusive comments. Equation 33 provides the formula for precision.
 \begin{equation}
\text { Precision }=\frac{\left|L_{T P}\right|}{\left|L_{T P}\right|+\left|L_{F P}\right|}
\end{equation}
Recall is determined by dividing the number of elements in the set of correctly predicted abusive labels $L_{TP}$  by the sum of number of elements in $L_{TP}$ set and falsely predicted non-abusive set $L_{FN}$. Formula for Recall is shown in Equation 34.
\begin{equation}
\text { Recall }=\frac{\left|L_{T P}\right|}{\left|L_{T p}\right|+\left|L_{F N}\right|}
\end{equation}

The F1-score metric, considers both precision and recall. It is determined as the harmonic mean of precision and recall. F1-score can be calculated as shown in Equation 35.

\begin{equation}
F 1-\text { score }=\frac{2 \times \text { precision } \times \text { recall }}{\text { precision }+\text { recall }}
\end{equation}

The accuracy metric is determined by dividing the total number of correctly predicted comments, be it abusive or non-abusive, by the total number of comments in the dataset. Since our dataset is fairly balanced, we use accuracy too for evaluation. Formula for Accuracy is given in Equation 36.
\begin{equation}
\text { Accuracy }=\frac{\left|L_{T P}\right|+\left|L_{T N}\right|}{\left|L_{T p}\right|+\left|L_{F N}\right|+\left|L_{T N}\right|+\left|L_{F P}\right|}
\end{equation}

Next, we discuss experimental results on both datasets using aforementioned evaluation metrics.
\vspace{-1mm}
\subsection {\textit{Experimental Results}} \label{label:experiment_results}

In this section, we present experimental results and comparisons. 
First, results of the proposed method are compared with classical and baseline state-of-the-art methods. Next, an ablation analysis is shown using different features with transliterated comments. Finally, language wise comparison is shown with state-of-the-art methods.

\subsubsection{\textit{Comparison with Classical and State-of-the-art Baseline Methods on SCIDN Dataset}}
To validate the efficacy of our method, we compare results of the proposed method with the existing methods. Comparison results on SCIDN dataset are shown in Table-\ref{table:compare_scidn}.

\begin{table}[h!]
\caption{Comparison with classical and state-of-the-art baseline methods on SCIDN dataset}
\centering
\resizebox{\textwidth}{!}{\begin{tabular}{llc c c c}
\toprule
\multicolumn{1}{c}{} &\textbf{Methods}                                                                 & \textbf{Acc}     & \textbf{P}    & \textbf{R}       & \textbf{F1}           \\ \midrule
\multicolumn{1}{c}{\multirow{3}{*}{Classical Methods}} &  Count-based{\citep{greevy2004classifying}}                     & 89.15                 & 91.92                 & 85.64                 & 88.67                 \\
\multicolumn{1}{c}{}                                               & Frequency-based{\citep{davidson2017automated}}                 & 89.18                 & 91.60                 & 86.16                 & 88.79                 \\
\multicolumn{1}{c}{}                                               & Embedding-based{\citep{mikolov2013efficient}}                 & 90.26                 & 93.28                 & 86.68                 & 89.86                 \\
 \midrule
\multicolumn{1}{c}{\multirow{7}{*}{State-of-the-art Baseline Methods}}              & 
                                              
Muril{\citep{khanuja2021muril}}      & 92.05   & 91.09      & 93.13      & 92.10     \\

\multicolumn{1}{c}{}                                               & M-Bert{\citep{devlin2019bert}}                          & 90.40                 & 91.06                 & 89.62                 & 90.34                 \\
\multicolumn{1}{c}{}                                               & XLM-R{\citep{conneau2020unsupervised}}                           & 91.64                 & 90.97                 & 92.47                 & 91.71                 \\
\multicolumn{1}{c}{}                                               & Modha \textit{et al.}{\citep{modha2020detecting}}             & \multicolumn{1}{l}{92.21} & \multicolumn{1}{l}{93.29} & \multicolumn{1}{l}{90.88} & \multicolumn{1}{l}{92.07} \\

\multicolumn{1}{c}{}                                               & Sazzed \textit{et al.}{\citep{sazzed2021abusive}} & 89.86                     & 92.12                     & 87.07                     & 89.53                 \\
\multicolumn{1}{c}{}                                               & Bansal \textit{et al.}{\citep{bansal2022transformer}} & {91.97}                     & {91.08}            & 92.98            & 92.02    \\ \midrule
\multicolumn{1}{c}{} & \textbf{Proposed Method }                                                                             & \textbf{95.40}        & \textbf{95.18}        & \textbf{95.59}        & \textbf{95.38}        \\ \bottomrule
\end{tabular}}
\label{table:compare_scidn}
\end{table}

It can be seen in the table that our method obtains an accuracy score of 95.40\% , precision of 95.18\%, recall of 95.59\% and F1-score of 95.38\%. Our method outperforms classical as well as state-of-the-art baselines by significant margins. The proposed method achieves a performance gain of 6.25\%, 3.25\%, 9.95\%, and 6.7\% over the count-based method, and 6.22\%, 3.58\%, 9.43\%, and 6.58\% over the frequency-based method in terms of accuracy, precision, recall and F1-score, respectively. This is due to the fact that count-based and frequency-based methods create text features based on the count of occurrence of each word and frequency of each word in the corpus, respectively. Such text features do not consider the context and similarity of words. On the other hand, we use contextual text features which consider both context and similarity of words. Besides these baselines, we also take non-contextual word-embeddings for comparison with our method. These word-embeddings are trained on the SCIDN dataset during fitting of the model. Our method performs significantly better than non-textual word-embedding based method by attaining 5.14\%, 1.9\%, 8.9\%, and 5.51\% higher accuracy, precision, recall and F1-score, respectively. 
Transformer-based models such as Muril, M-Bert and XLM-R have proved to be very effective for multilingual text classification problems. The major advantage of these methods is that they provide contextual embeddings. Our approach capitalizes the benefits of these transformer-based methods and also includes post tendency and user history. Our method outperforms Muril by 3.35\%, 4.09\%, 2.46\%, and 3.28\%, M-Bert by 5\%, 4.12\%, 5.96\%, and 5.03\% in terms of accuracy, precision, recall and F1-score, respectively. XLM-R is beaten by 3.76\%, 4.21\%, 3.12\%, and 3.67\%  in accuracy, precision, recall and F1-score metrics, respectively.
 We compare our results with Modha \textit{et al.} \citep{modha2020detecting} which uses an embedding layer with an attention mechanism. Their neural network contains two layers of LSTM as forward and backward layers. Our method achieves a performance gain of 3.19\%, 1.89\%, 4.71\% and 3.31\% over Modha \textit{et al.} in terms of accuracy, precision, recall and F1-score, respectively. 
A significant performance gain across all four evaluation metrics establishes the effectiveness of our proposed method. We also compare results of our method with Sazzed \textit{et al.}\citep{sazzed2021abusive} which employs SVM with TF-IDF. They remove rows with words written in native script. They do not deal with the code-mixed aspect. We, on the other hand, use cross-lingually trained word-embeddings in the ensemble framework to deal with code-mixing, and a large set of abusive words in 12 Indic languages to deal with spelling errors. Our method outperforms Sazzed \textit{et al.} by 5.54\%, 3.06\%, 8.52\%, and 5.85\% in accuracy, precision, recall and F1-score metrics, respectively. On comparison with Bansal \textit{et al.}~\citep{bansal2022transformer}, our method achieves an increase of 3.43\%, 4.1\%, 2.61\%, and 3.36\% in terms of accuracy, precision, recall and F1-score, respectively. Bansal \textit{et al.} uses XLM-R with transliterated comments and emoji embeddings. 

\subsubsection{\textit{Comparison with Classical and State-of-the-art Baseline Methods on MACI Dataset}}
We compare the results of our method on MACI with the same set of previously mentioned methods. We use post-polarity for MACI instead of the combined user-post polarity feature due to an unavailability of user features in this dataset. On this dataset as well, our method exhibits a significantly better performance compared to other methods. This is due to the reason that the post polarity feature has a strong association with abusive content. All the results are shown in Table-\ref{table:compare_maci}.
As shown in the table, our method achieves 91.97\% accuracy, 90.12\% precision, 93.08\% recall and an F1-score of 91.58\%. We achieve an absolute performance gain of 10.21\%, 4.52\%, 19.37\%, and 12.37\% over the count-based method, and 10.08\%, 3.42\%, 20.37\%, and 12.49\% over the frequency-based method in terms of accuracy, precision, recall and F1-score, respectively. Another baseline, the embedding-based method, which refers to non-contextual word-embeddings, is outperformed too by our method on the MACI dataset. Our method obtains 7.54\%, 3.07\%, 14.31\% and 8.88\% higher accuracy, precision, recall and F1-score, respectively. The transformer-based method Muril is outperformed by 9\%, 2.95\%, 17.98\% and 10.9\%, M-Bert by 8.5\%, 0.2\%, 19.78\%, and 10.81\%, XLM-R by 6.13\%, 4.96\%, 8.18\%, and 6.56\%, in terms of accuracy, precision, recall and F1-score, respectively. Besides these, results are compared with the existing methods Modha \textit{et al.}~\citep{modha2020detecting}, Sazzed \textit{et al.}~\citep{sazzed2021abusive} and Bansal \textit{et al.}~\citep{bansal2022transformer}. Our method achieves a performance gain of 10.12\%, 5.52\%, 18.15\%, and 12.11\% over Modha \textit{et al.} in terms of accuracy, precision, recall and F1-score evaluation metrics, respectively. On comparison with Sazzed \textit{et al.}, or method achieves an increase of 9.66\%, 3.96\%, 18.46\%, and 11.6\% in accuracy, precision, recall and F1-score, respectively. Bansal \textit{et al.} performed their experiments on the same dataset, our proposed method outperforms their method by 4.27\%, 2.42\%, 7.28\%, and 5.18\% in in terms of accuracy, precision, recall and F1-score evaluation metrics, respectively.

\begin{table}[h!]
\caption{Comparison with classical and state-of-the-art baseline methods on MACI dataset}
\centering
\resizebox{\textwidth}{!}{\begin{tabular}{llc c c c}
\toprule
\multicolumn{1}{c}{} &\textbf{Methods}                                                                & \textbf{Acc} & \textbf{P} & \textbf{R} & \textbf{F1}    \\ \midrule
\multicolumn{1}{c}{\multirow{3}{*}{Classical Methods }} & Count-based{\citep{greevy2004classifying}}                    & 81.76             & 85.60              & 73.71           & 79.21          \\
\multicolumn{1}{c}{}                                               & Frequency-based{\citep{davidson2017automated}}                & 81.89             & 86.70              & 72.70           & 79.09          \\
\multicolumn{1}{c}{}                                               & Embedding-based{\citep{mikolov2013efficient}}                & 84.43             & 87.04              & 78.77           & 82.70          \\
 \midrule
\multicolumn{1}{c}{\multirow{5}{*}{State-of-the-art Baseline Methods}}              &

Muril{\citep{khanuja2021muril}}                          & 82.97             & 87.17              & 75.09           & 80.68          \\  
\multicolumn{1}{c}{}                                               & M-Bert{\citep{devlin2019bert}}                         & 83.47             & 89.92              & 73.30           & 80.77          \\
\multicolumn{1}{c}{}                                               & XLM-R{\citep{conneau2020unsupervised}}                          & 85.84             & 85.15              & 84.89           & 85.02          \\

\multicolumn{1}{c}{}                                    & Modha \textit{et al.}{\citep{modha2020detecting}]}            & \multicolumn{1}{l}{81.85} & \multicolumn{1}{l}{84.60} & \multicolumn{1}{l}{74.92} & \multicolumn{1}{l}{79.47} \\

\multicolumn{1}{c}{}                                               & Sazzed \textit{et al.}{\citep{sazzed2021abusive}} & 82.31                     & 86.16                     & 74.62                     & 79.98                 \\
\multicolumn{1}{c}{}                                               & Bansal \textit{et al.}{\citep{bansal2022transformer}} & 87.70                     &  87.70           & 85.80             & 86.40        \\ \midrule
\multicolumn{1}{c}{} & \textbf{Proposed Method}                                                                            & \textbf{91.97}    & \textbf{90.12}     & \textbf{93.08}  & \textbf{91.58} \\ \bottomrule
\end{tabular}}
\label{table:compare_maci}
\end{table}

Our proposed approach has shown significant improvement over other methods on both the datasets. Our proposed method makes predictions based on two aspects of the content, i.e., social context features including user history and post polarity, and textual features. Results presented in Table-\ref{table:compare_scidn} and Table-\ref{table:compare_maci} show that the utilization of user history and post polarity improves the performance of our proposed method remarkably.

\subsubsection{\textit{Ablation Analysis on SCIDN Dataset}}
In this section, we discuss and contrast the results utilizing different features with preprocessed transliterated comments while performing the experiments on the SCIDN dataset.
\begin{figure}[h!]
    \centering
    \includegraphics[width=12cm, height=7cm]{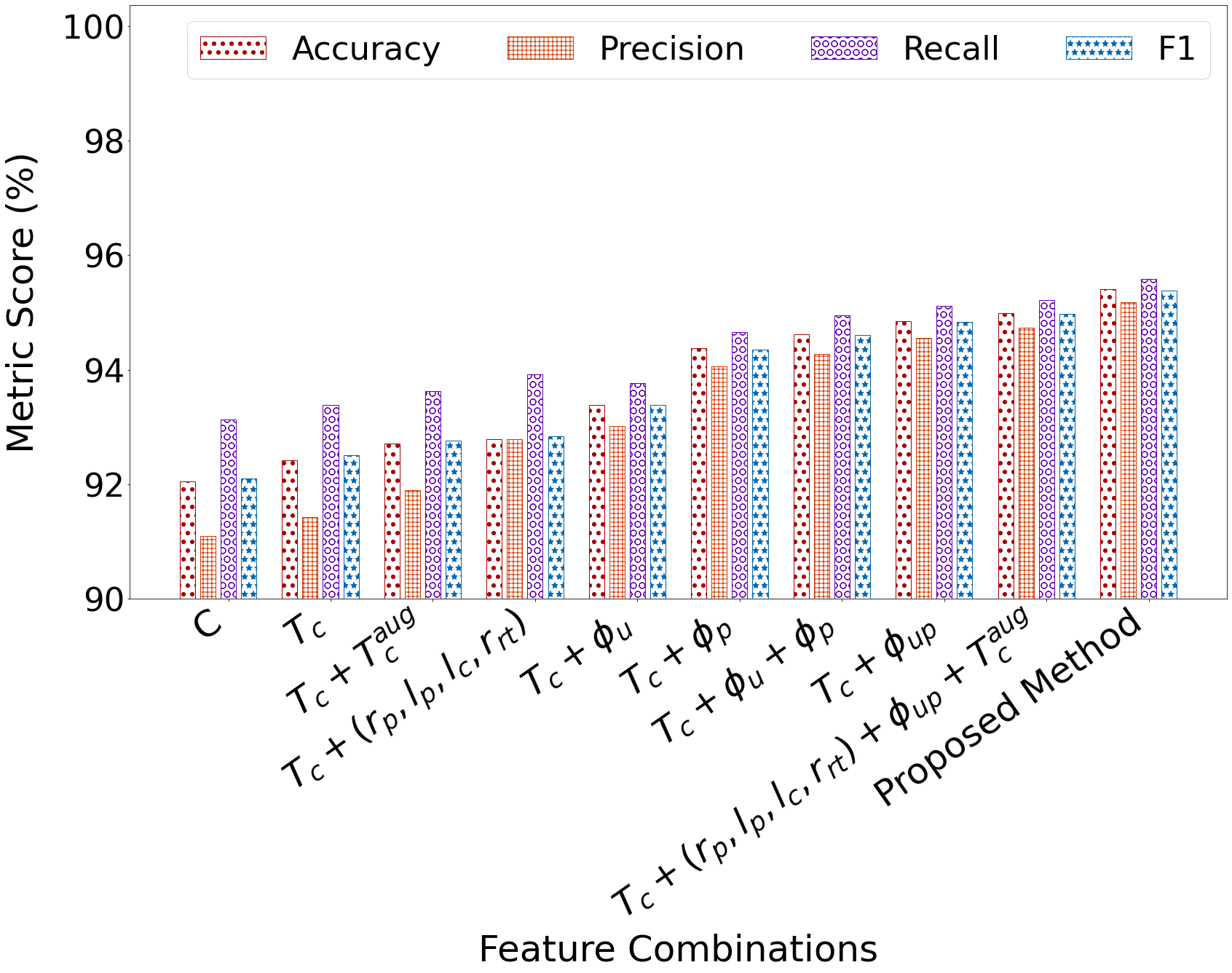}
    \caption{Ablation analysis results on SCIDN dataset}
    \label{fig:ablation_raw_scidn}
\end{figure}
Overall, our proposed model achieves an accuracy of 95.40\%, precision of 95.18\%, recall of 95.59\% and an F1 score of 95.38\%, which is 3.35\%, 4.09\%, 2.46\% and 3.28\%  higher respectively, in comparison to the results taken on raw comments. Figure-\ref{fig:ablation_raw_scidn} depicts the graph which compares the results for different feature combinations for the SCIDN dataset.\\
For the ablation study, we first take results with raw comments using their contextual embeddings from the transformer-based model, which gives an accuracy of 92.05\%, precision of 91.09\%, recall of 93.13\%, and an F1-score of 92.10\%.
We then perform the experiments with preprocessed transliterated comments using their contextual embeddings, and here we obtain 0.37\%, 0.34\%, 0.26\% and 0.41\% better results than the results of raw comments in terms of accuracy, precision, recall and F1-score, respectively. We then take results after data augmentation, and here also we get a slight improvement in performance as compared to raw comments.
Next, along with the contextual embeddings from the transliterated comments, we use the four post features and through them, an improvement of 0.74\% , 0.68\%, 0.79\% and 0.74\% is observed in accuracy, precision, recall and F1-score evaluation metrics, respectively.
It can be seen in the graph that the highest improvement in results, by using an individual feature  along with the embeddings, is achieved by combined user-post polarity as compared to the results of raw comments. Here, we get significantly higher accuracy, precision, recall and F1-score of 94.85\%, 94.56\%, 95.11\% and 94.83\%, respectively. These results are 2.79\%, 3.47\%, 1.98\%, and 2.73\% better than the results on raw comments in terms of accuracy, precision, recall and F1-score, respectively. This performance gain is expected because user-post polarity has the strongest correlation with class labels as shown in Table-\ref{table:point_scidn} in Section-\ref{section:social_context_feature}.
Each feature contributes in performance gain when used in combination with the contextual embeddings, thus, finally providing state-of-the-art results as compared to the results for single modality, i.e text. Social context features, specifically user and post polarities, are recommended for abusive content detection.

\begin{figure}[h!]
    \centering
    \includegraphics[width=12cm, height=7cm]{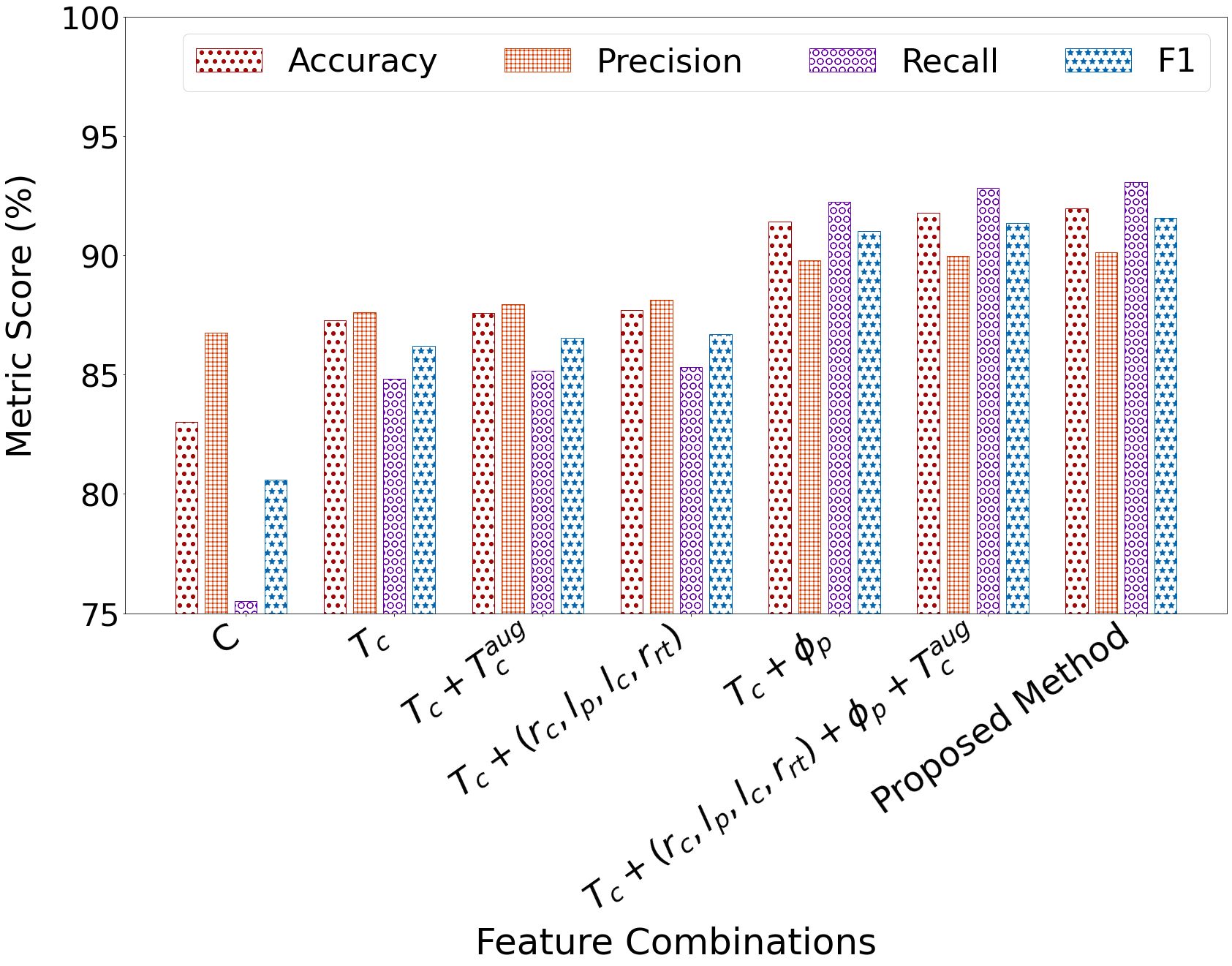}
    \caption{Ablation anlaysis results on MACI dataset}
    \label{fig:ablation_raw_maci}
\end{figure}

\subsubsection{\textit{Ablation Analysis on MACI Dataset}}
In this section, an ablation analysis is presented on the MACI dataset. Figure-\ref{fig:ablation_raw_maci} depicts the graph which compares the results for different feature combinations for the MACI dataset.
In this analysis, we discuss and contrast the results of preprocessed transliterated comments with raw comments by using different features while performing the experiments.
Using all features, our proposed model achieves an accuracy of 91.97\% , precision of 90.12\%, recall of 93.08\%, and an F1-score of 91.58\%, which is 8.96\%, 4.37\%, 17.56\% and 10.98\% higher respectively, in comparison to the results for raw comments.
For the MACI dataset, post polarity provides the best results for any individual feature used in combination with text embeddings. This is because post polarity has the strongest correlation with class labels as shown in Table-\ref{table:point_maci} in Section-\ref{section:social_context_feature}. Using post polarity, we achieve 91.43\%, 89.78\%, 92.26\%, and 91.01\% accuracy, precision, recall and F1-score, respectively. This is an improvement of 8.42\%, 3.03\%, 16.74\%, and 10.41\%  over raw comments in terms of respective evaluation metrics.

\subsubsection{\textit{Language-wise Performance on SCIDN}} We perform a language-wise comparison to validate the performance of our proposed method for different Indic languages. Our method outperforms all the compared methods with substantial margins. We use transformer-based multilingual methods for comparison.

\begin{table}[h!]
\caption{Language-wise comparison on SCIDN dataset}
\scriptsize
\centering
\resizebox{\textwidth}{!}{\begin{tabular}{cllll llll llll lll}
\toprule
\multirow{2}{*}{\textbf{Language}} & \multicolumn{3}{c}{\textbf{Muril}} &\multicolumn{1}{l}{}                             & \multicolumn{3}{c}{\textbf{XLM-R}} &\multicolumn{1}{l}{}                            & \multicolumn{3}{c}{\textbf{M-Bert}} &\multicolumn{1}{l}{}                           & \multicolumn{3}{c}{\textbf{Proposed Method }}                              \\ \cmidrule(lr){2-4} \cmidrule(lr){6-8}\cmidrule(lr){10-12} \cmidrule(lr){14-16}
                                   & \multicolumn{1}{c}{\textbf{P}}     & \multicolumn{1}{c}{\textbf{R}}     & \multicolumn{1}{c}{\textbf{F1}} &\multicolumn{1}{l}{}  &\multicolumn{1}{c}{\textbf{P}}      & \multicolumn{1}{c}{\textbf{R}}     & \multicolumn{1}{c}{\textbf{F1}} &\multicolumn{1}{l}{}    & \multicolumn{1}{c}{\textbf{P}}     & \multicolumn{1}{c}{\textbf{R}}     & \multicolumn{1}{c}{\textbf{F1}} &\multicolumn{1}{l}{}    & \multicolumn{1}{c}{\textbf{P}}     & \multicolumn{1}{c}{\textbf{R}}     & \multicolumn{1}{c}{\textbf{F1}}    \\ \midrule
As                                 & \multicolumn{1}{l}{81.30}  & \multicolumn{1}{l}{88.77} & 84.87 &\multicolumn{1}{l}{} & \multicolumn{1}{l}{90.56} & \multicolumn{1}{l}{88.88} & 89.72 &\multicolumn{1}{l}{} & \multicolumn{1}{l}{88.46} & \multicolumn{1}{l}{85.18} & 86.79 &\multicolumn{1}{l}{} & \multicolumn{1}{l}{{94.00}}    & \multicolumn{1}{l}{{95.91}} & {94.94} \\
Bn                                 & \multicolumn{1}{l}{89.52} & \multicolumn{1}{l}{91.59} & 90.54&\multicolumn{1}{l}{} & \multicolumn{1}{l}{88.39} & \multicolumn{1}{l}{92.27} & 90.29 &\multicolumn{1}{l}{} & \multicolumn{1}{l}{89.53} & \multicolumn{1}{l}{83.85} & 86.59 &\multicolumn{1}{l}{} & \multicolumn{1}{l}{{92.46}} & \multicolumn{1}{l}{{92.58}} & {92.52} \\ 
En                                 & \multicolumn{1}{l}{90.20}  & \multicolumn{1}{l}{93.39} & 91.77 &\multicolumn{1}{l}{} & \multicolumn{1}{l}{91.43} & \multicolumn{1}{l}{93.39} & 92.40 &\multicolumn{1}{l}{}  & \multicolumn{1}{l}{92.00}    & \multicolumn{1}{l}{92.66} & 92.33 &\multicolumn{1}{l}{} & \multicolumn{1}{l}{{95.46}} & \multicolumn{1}{l}{{95.77}} & {95.61} \\ 
Gu                                 & \multicolumn{1}{l}{89.49} & \multicolumn{1}{l}{90.70}  & 90.09 &\multicolumn{1}{l}{} & \multicolumn{1}{l}{87.50}  & \multicolumn{1}{l}{92.98} & 90.15 &\multicolumn{1}{l}{} & \multicolumn{1}{l}{86.20}  & \multicolumn{1}{l}{87.71} & 86.95 &\multicolumn{1}{l}{} & \multicolumn{1}{l}{{93.86}} & \multicolumn{1}{l}{{93.43}} & {93.65} \\ 
Hi                                 & \multicolumn{1}{l}{92.44} & \multicolumn{1}{l}{94.19} & 93.31 &\multicolumn{1}{l}{} & \multicolumn{1}{l}{93.11} & \multicolumn{1}{l}{93.34} & 93.22 &\multicolumn{1}{l}{} & \multicolumn{1}{l}{92.73} & \multicolumn{1}{l}{92.14} & 92.43 &\multicolumn{1}{l}{} & \multicolumn{1}{l}{{96.39}} & \multicolumn{1}{l}{{97.25}} & {96.82} \\ 
Ka                                 & \multicolumn{1}{l}{84.87} & \multicolumn{1}{l}{88.84} & 86.81 &\multicolumn{1}{l}{} & \multicolumn{1}{l}{87.89} & \multicolumn{1}{l}{85.48} & 86.62 &\multicolumn{1}{l}{} & \multicolumn{1}{l}{88.33} & \multicolumn{1}{l}{80.76} & 84.38 &\multicolumn{1}{l}{} & \multicolumn{1}{l}{{95.49}} & \multicolumn{1}{l}{{88.51}} & {91.87} \\ 
Ml                                 & \multicolumn{1}{l}{79.27} & \multicolumn{1}{l}{90.39} & 84.47 &\multicolumn{1}{l}{} & \multicolumn{1}{l}{73.52} & \multicolumn{1}{l}{75.48} & 74.49 &\multicolumn{1}{l}{} & \multicolumn{1}{l}{78.30}  & \multicolumn{1}{l}{51.69} & 62.28 &\multicolumn{1}{l}{} & \multicolumn{1}{l}{{95.41}} & \multicolumn{1}{l}{{89.46}} & {92.34} \\ 
Mr                                 & \multicolumn{1}{l}{86.47} & \multicolumn{1}{l}{95.14} & 90.59 &\multicolumn{1}{l}{} & \multicolumn{1}{l}{84.92} & \multicolumn{1}{l}{94.66} & 89.53 &\multicolumn{1}{l}{} & \multicolumn{1}{l}{84.17} & \multicolumn{1}{l}{88.67} & 86.36 &\multicolumn{1}{l}{} & \multicolumn{1}{l}{{94.80}}  & \multicolumn{1}{l}{{96.80}}  & {95.79} \\ 
Od                                 & \multicolumn{1}{l}{90.60}  & \multicolumn{1}{l}{92.16} & 91.37 &\multicolumn{1}{l}{} & \multicolumn{1}{l}{91.90}  & \multicolumn{1}{l}{90.32} & 91.11 &\multicolumn{1}{l}{} & \multicolumn{1}{l}{92.43} & \multicolumn{1}{l}{89.54} & 90.96 &\multicolumn{1}{l}{} & \multicolumn{1}{l}{{95.70}}  & \multicolumn{1}{l}{{92.53}} & {93.95} \\ 
Pa                                 & \multicolumn{1}{l}{93.16} & \multicolumn{1}{l}{95.96} & 94.54 &\multicolumn{1}{l}{} & \multicolumn{1}{l}{92.57} & \multicolumn{1}{l}{96.01} & 94.26 &\multicolumn{1}{l}{} & \multicolumn{1}{l}{92.21} & \multicolumn{1}{l}{94.93} & 93.55 &\multicolumn{1}{l}{} & \multicolumn{1}{l}{{94.30}}  & \multicolumn{1}{l}{{96.83}} & {95.55} \\ 
Ta                                 & \multicolumn{1}{l}{86.88} & \multicolumn{1}{l}{85.13} & 86.00  &\multicolumn{1}{l}{}  & \multicolumn{1}{l}{83.83} & \multicolumn{1}{l}{82.75} & 83.29 &\multicolumn{1}{l}{} & \multicolumn{1}{l}{84.27} & \multicolumn{1}{l}{67.92} & 75.22 &\multicolumn{1}{l}{} & \multicolumn{1}{l}{{91.84}} & \multicolumn{1}{l}{{89.94}} & {90.88} \\ 
Te                                 & \multicolumn{1}{l}{87.89} & \multicolumn{1}{l}{94.04} & 90.86 &\multicolumn{1}{l}{} & \multicolumn{1}{l}{87.84} & \multicolumn{1}{l}{92.61} & 90.16 &\multicolumn{1}{l}{} & \multicolumn{1}{l}{87.51} & \multicolumn{1}{l}{89.89} & 88.69 &\multicolumn{1}{l}{} & \multicolumn{1}{l}{{95.74}} & \multicolumn{1}{l}{{93.68}} & {94.70}  \\ \hline
\end{tabular}}
\label{table:scidn_lang}
\end{table}

Results for language-wise comparison corresponding to SCIDN are presented in Table-\ref{table:scidn_lang}. Our method outperforms Muril, XLM-R and M-Bert for Assamese by 10.07\%,  5.22\%,  and  8.15\%, respectively, in terms of F1-score. For the Bengali language, our model achieves better F1-scores as compared to the results of the aforementioned models by 1.98\%,  2.23\%,  and  5.93\%, respectively, in terms of the F1-score metric. Corresponding to the same models, we have achieved an improvement through our own model, of  3.84\%,  3.21\%,  and  3.28\%, for English. An improvement of 3.56\%,  3.50\%,  and  6.70\%, is observed for Gujarati, of 3.51\%,  3.6\%,  and  4.39\%, for Hindi, and of 5.06\%,  5.25\%,  and  7.49\%, for Kannada, in terms of F1-score. For Malayalam, our model has improved abuse identification by 7.87\%,  17.85\%,  and  30.06\%, in comparison to the state-of-the-art models referenced above, by 5.20\%,  6.26\%,  and  9.43\%, for Marathi, and by 2.58\%,  2.84\%,  \&  2.99\%, for Odiya, for the F1-score metric. For Punjabi, the margin of performance gain is of 1.01\%,  1.29\%,  and  2\%, for Tamil it is  4.88\%,  7.59\%,  and  15.66\% , and for Telugu it is 3.84\%,  4.54\%, and  6.01\% in terms of F1-score. For precision and recall too, our method outperforms all the compared methods with significant margins as depicted in the table.

\subsubsection{\textit{Language-wise Performance on MACI}} In this section, a language-wise performance comparison is presented for the MACI dataset. Comparisons are made with tranformer-based methods, Muril, M-Bert and XLM-R. For the MACI dataset too, our method outperforms all the compared methods for all languages. Table-\ref{table:maci_lang} shows results of language-wise comparisons on the MACI dataset. 
In Assamese, our model beats the performance of Muril, XLM-R and M-Bert by 13.07\%, 15.02\% and 15.6\%, respectively, in terms of F1-score. Our model has achieved a performance gain of 10.25\%, 5.81\%, and 9.53\% in F1-score for the Bengali language, as compared to the aforementioned models. In Gujarati, our approach outperforms these state-of-the-art models by 13.9\%, 6.84\% and 8.25\%, in Hindi by 9.91\%, 8.76\%, and 11.61\%, and in Kannada by 14.49\%, 7.34\%, and 9.32\%, respectively, in terms of F1-score. The experimentation we conducted has shown that for Malayalam, our model is better than the aforementioned models by 55.92\%, 23.20\%, and 62.77\%, while for Marathi, it is better than them by 22.56\%, 13.44\%, and 29.27\%, respectively, for the F1-score metric. We achieve an improvement in F1-score of 13.92\%, 5.73\%, and 12.97\%, for Odiya, in comparison to other models and for Tamil and Telugu, we are outperforming these models by 18.23\%, 3.44\%, and 7.82\%, and 7.66\%, 6.06\%, and 7.22\%, respectively, in terms of F1-score. Our model outperforms Muril, XLM-R and M-Bert in language-wise comparisons pertaining to the precision and recall metrics as well, as depicted in the table.

\begin{table}[h!]
\scriptsize
\centering
\caption{Language-wise comparison on MACI dataset}
\resizebox{\textwidth}{!}{\begin{tabular}{cllll llll llll lll}
\toprule
\multirow{2}{*}{\textbf{Language}} & \multicolumn{3}{c}{\textbf{Muril}} &\multicolumn{1}{l}{}                             & \multicolumn{3}{c}{\textbf{XLM-R}} &\multicolumn{1}{l}{}                            & \multicolumn{3}{c}{\textbf{M-Bert}} &\multicolumn{1}{l}{}                           & \multicolumn{3}{c}{\textbf{Proposed Method }}                              \\ \cmidrule(lr){2-4} \cmidrule(lr){6-8}\cmidrule(lr){10-12} \cmidrule(lr){14-16}
                                   & \multicolumn{1}{c}{\textbf{P}}     & \multicolumn{1}{c}{\textbf{R}}     & \multicolumn{1}{c}{\textbf{F1}} &\multicolumn{1}{l}{}  &\multicolumn{1}{c}{\textbf{P}}      & \multicolumn{1}{c}{\textbf{R}}     & \multicolumn{1}{c}{\textbf{F1}} &\multicolumn{1}{l}{}    & \multicolumn{1}{c}{\textbf{P}}     & \multicolumn{1}{c}{\textbf{R}}     & \multicolumn{1}{c}{\textbf{F1}} &\multicolumn{1}{l}{}    & \multicolumn{1}{c}{\textbf{P}}     & \multicolumn{1}{c}{\textbf{R}}     & \multicolumn{1}{c}{\textbf{F1}}    \\ \midrule
As                        & \multicolumn{1}{l}{87.65}          & \multicolumn{1}{l}{80.68}          & 84.23  &\multicolumn{1}{l}{}        & \multicolumn{1}{l}{82.75}          & \multicolumn{1}{l}{81.81} & 82.28 &\multicolumn{1}{l}{} & \multicolumn{1}{l}{88.15} & \multicolumn{1}{l}{76.13} & 81.70 &\multicolumn{1}{l}{} & \multicolumn{1}{l}{{96.78}} & \multicolumn{1}{l}{{97.83}} & {97.30} \\
Bn                                 & \multicolumn{1}{l}{85.31}          & \multicolumn{1}{l}{86.57}          & 85.94 &\multicolumn{1}{l}{}          & \multicolumn{1}{l}{89.89}          & \multicolumn{1}{l}{90.88} & 90.38 &\multicolumn{1}{l}{} & \multicolumn{1}{l}{92.33} & \multicolumn{1}{l}{81.65} & 86.66 &\multicolumn{1}{l}{} & \multicolumn{1}{l}{{94.58}} & \multicolumn{1}{l}{{97.86}} & {96.19} \\
Gu                                 & \multicolumn{1}{l}{77.28}          & \multicolumn{1}{l}{86.18}          & 81.49   &\multicolumn{1}{l}{}       & \multicolumn{1}{l}{89.29}          & \multicolumn{1}{l}{87.82} & 88.55 &\multicolumn{1}{l}{} & \multicolumn{1}{l}{88.47} & \multicolumn{1}{l}{85.85} & 87.14 &\multicolumn{1}{l}{} & \multicolumn{1}{l}{{93.29}} & \multicolumn{1}{l}{{97.59}} & {95.39} \\
Hi                                 & \multicolumn{1}{l}{88.35}          & \multicolumn{1}{l}{81.94}          & 85.02  &\multicolumn{1}{l}{}        & \multicolumn{1}{l}{84.76}          & \multicolumn{1}{l}{87.63} & 86.17 &\multicolumn{1}{l}{} & \multicolumn{1}{l}{89.58} & \multicolumn{1}{l}{77.87} & 83.32 &\multicolumn{1}{l}{} & \multicolumn{1}{l}{{92.57}} & \multicolumn{1}{l}{{97.42}} & {94.93} \\
Ka                                 & \multicolumn{1}{l}{89.05}          & \multicolumn{1}{l}{75.62}          & 81.78  &\multicolumn{1}{l}{}        & \multicolumn{1}{l}{87.42}          & \multicolumn{1}{l}{90.49} & 88.93 &\multicolumn{1}{l}{} & \multicolumn{1}{l}{87.13} & \multicolumn{1}{l}{86.77} & 86.95 &\multicolumn{1}{l}{} & \multicolumn{1}{l}{{95.58}} & \multicolumn{1}{l}{{96.96}} & {96.27} \\
Ml                                 & \multicolumn{1}{l}{71.49}          & \multicolumn{1}{l}{23.79}          & 35.70    &\multicolumn{1}{l}{}      & \multicolumn{1}{l}{72.80}          & \multicolumn{1}{l}{64.54} & 68.42 &\multicolumn{1}{l}{} & \multicolumn{1}{l}{72.04} & \multicolumn{1}{l}{18.04} & 28.85 &\multicolumn{1}{l}{} & \multicolumn{1}{l}{{92.47}} & \multicolumn{1}{l}{{90.78}} & {91.62} \\
Mr                                 & \multicolumn{1}{l}{76.69}          & \multicolumn{1}{l}{54.01}          & 63.38    &\multicolumn{1}{l}{}      & \multicolumn{1}{l}{77.26}          & \multicolumn{1}{l}{68.82} & 72.80 &\multicolumn{1}{l}{} & \multicolumn{1}{l}{85.08} & \multicolumn{1}{l}{42.82} & 56.97 &\multicolumn{1}{l}{} & \multicolumn{1}{l}{{87.60}} & \multicolumn{1}{l}{{84.92}} & {86.24} \\
Od                                 & \multicolumn{1}{l}{82.88}          & \multicolumn{1}{l}{72.06}          & 77.09   &\multicolumn{1}{l}{}       & \multicolumn{1}{l}{82.96}          & \multicolumn{1}{l}{87.72} & 85.28 &\multicolumn{1}{l}{} & \multicolumn{1}{l}{84.03} & \multicolumn{1}{l}{72.84} & 78.04 &\multicolumn{1}{l}{} & \multicolumn{1}{l}{{89.47}} & \multicolumn{1}{l}{{92.61}} & {91.01} \\
Ta                                 & \multicolumn{1}{l}{{93.67}} & \multicolumn{1}{l}{61.63} & {74.34} &\multicolumn{1}{l}{} & \multicolumn{1}{l}{91.03} & \multicolumn{1}{l}{87.31} & 89.13 &\multicolumn{1}{l}{} & \multicolumn{1}{l}{91.64} & \multicolumn{1}{l}{78.83} & 84.75 &\multicolumn{1}{l}{} & \multicolumn{1}{l}{90.05} & \multicolumn{1}{l}{{95.24}} & {92.57} \\
Te                                 & \multicolumn{1}{l}{87.38}          & \multicolumn{1}{l}{81.22}          & 84.19  &\multicolumn{1}{l}{}        & \multicolumn{1}{l}{86.93}          & \multicolumn{1}{l}{84.76} & 85.79 &\multicolumn{1}{l}{} & \multicolumn{1}{l}{93.21} & \multicolumn{1}{l}{77.50} & 84.63 &\multicolumn{1}{l}{} & \multicolumn{1}{l}{{88.89}} & \multicolumn{1}{l}{{95.01}} & {91.85} \\ \hline
\end{tabular}}
\label{table:maci_lang}
\end{table}

\section{Discussion}
\label{sec:sa_dscn}
This article introduces a technique for abusive content detection on social media. Proposed method is specifically built for multiple low-resource Indic languages. Unlike recent trend to focus solely on textual characteristics to classify the content, our method integrates social context features with textual features. Integration of features is done using late fusion technique, where two categories of features are fed to two separate networks before concatenating them to get the joint feature for final classification. Effectiveness of the features is first shown using point biserial correlation in Table-\ref{table:point_scidn} and Table-\ref{table:point_maci}, and finally in the ablation study section using graphical representation in Figure-\ref{fig:ablation_raw_scidn} and \ref{fig:ablation_raw_maci}. The proposed method targets the misspelled abusive words by learning multliple spellings of such words. To this end, a comprehensive set of abusive words is used for data augmentation. This set contains multiple spelling variations of each abusive word, covering a large number prospective spellings which are introduced either by user made mistakes or during the transliteration of comments. To deal with uncertain grammatical structures of comments due to code-mixed text, we exploit cross-lingually trained transformer-based method. Overall comparison results show that proposed method outperforms the classical as well state-of-the-art methods by significant margin.
One of the limitations of the proposed method is that it requires user history; therefore, in the absence of a user's past comments, the proposed method may not be as effective as it is with the user history. 
Ensuring the model's resilience to such errors needs additional work beyond the scope of our work. 

\section{Conclusion}
\label{sec:sa_cnfw}
In this paper, we propose a novel multilingual abusive content detection method for Indic languages. Our method is not only text aware but also takes user history and post affinity into account. We recommend that user and post features can be utilized to determine user and post affinity towards abusive content. These affinities, along with comment text and post metadata, are used for classification in our proposed approach. We use contextual text embeddings as text features and employ a dedicated module to learn the text features. Another module is employed for social context feature learning. Joint representation of these learnt features, when used with a Deep Learning model, improves the performance by a significant margin. Further, we take cross-lingually trained text embeddings in the ensemble framework to obtain better predictive performance. We show with extensive testing on two datasets that our proposed method outperforms the baseline and state-of-the-art methods. Our model achieves better performance across all the evaluation metrics as established by the experimental results.

\bibliographystyle{apalike}

\bibliography{main}





\end{document}